\pdfoutput=1
\documentclass[11pt]{article}
\usepackage[margin=1in]{geometry}
\usepackage{amsmath,amssymb,amsthm,booktabs}
\usepackage[hidelinks]{hyperref}
\usepackage{tikz}
\usetikzlibrary{arrows.meta,backgrounds}
\tikzset{
  v/.style={circle,draw,fill=white,inner sep=0pt,minimum size=7pt,line width=0.6pt},
  d/.style={v,fill=black},
  e/.style={line width=0.6pt},
  faint/.style={line width=0.5pt,black!35},
  jn/.style={line width=1.6pt,-{Stealth[length=5pt]},shorten >=3.5pt},
  cut/.style={line width=0.6pt,dashed},
  wl/.style={font=\tiny,fill=white,inner sep=2pt},
  fcap/.style={font=\footnotesize,align=center},
  lab/.style={font=\scriptsize},
}
\def\exvertices{1/0/0,2/2/0.3,3/4/0,4/1/1.6,5/3/1.7,6/0/3,7/2/3.2,8/4/3}
\def\exedges{1/2/3,2/3/1,1/4/2,2/4/1,2/5/4,3/5/2,4/5/1,4/6/3,4/7/2,5/7/1,5/8/3,6/7/2,7/8/1,3/8/2}

\newtheorem{theorem}{Theorem}
\newtheorem{lemma}[theorem]{Lemma}
\newtheorem{corollary}[theorem]{Corollary}
\newtheorem{proposition}[theorem]{Proposition}
\theoremstyle{remark}

\newcommand{\E}{\mathbb{E}}
\newcommand{\cut}{\mathrm{cut}}

\title{Random Centres and the \texorpdfstring{$\varepsilon$}{epsilon}-Dependence of\\ Constant-Round Planar Domination}
\hypersetup{pdftitle={Random Centres and the epsilon-Dependence of Constant-Round Planar Domination},pdfauthor={Sebastian Senge}}
\author{Sebastian Senge\\ Independent Researcher}
\date{}

\begin{document}
\maketitle

\begin{abstract}
It was recently shown (arXiv:2610.09618) that the infimum of the approximation ratios of deterministic
constant-round LOCAL algorithms for minimum dominating set on planar graphs is~$7$. The
$(7+\varepsilon)$-approximation given there runs in $T_0+O((c_0/\varepsilon)^{\kappa})$ rounds with
$\kappa=\ln 3/\ln(24/23)\approx 25.81$, and it is asked there whether this exponent can be
substantially reduced. We replace the heaviest-edge star selection in its clustering step by a
random-centre rule. Every part becomes a centre independently with probability~$p$, and every
non-centre joins its heaviest centre neighbour. For planar graphs, a layer-cake count combined
with two degree inequalities shows that for $p\ge21/64$ each level contracts at least an expected
$(1-p)\big(3-2(1-p)^3-(1-p)^6\big)/9$ fraction of the remaining weight. With $p=3/8$, which takes three fair coins per part and level,
this gives $\kappa<5.89$. With $p=1/2$, that is, one fair coin per part and level as in the
original algorithm, it gives $\kappa<6.7$. An analysis that counts only these join edges cannot
give an exponent below $5.74$. For every proper minor-closed class of density $a$, a
bounded out-degree orientation gives the fraction $(1-p)(1-(1-p)^k)/k$ with $k=\lceil a\rceil$.
The new exponent is never worse, and with about $\frac12\log_2 a$ fair coins per part and level
it is smaller by a factor that tends to $8$ as the density grows. This is optimal for the rule when $a$ is an integer and the class contains all $K_{a,n}$.
A bipartite density bound or a non-integer density improves it further. In particular, the planar
bounds hold verbatim on the torus, the projective plane and the Klein bottle. We then turn to the
dependence on $\varepsilon$ of the $(1+\varepsilon)$-approximations. On paths, every deterministic
$r$-round algorithm for fractional domination, and every randomized one that always outputs a
dominating set, has (expected) ratio at least $1+1/(r+4)$. So these approximations need $\Omega(1/\varepsilon)$
rounds on every minor-closed class that contains all paths. For fractional domination on paths of
known length, this is tight up to an additive constant. On forests and on outerplanar graphs, with no bound on the degree,
$O(1/\varepsilon)$ rounds suffice for both approximations, so there the optimal number of rounds is
$\Theta(1/\varepsilon)$. Machine-checked in Lean~4 with Mathlib are the level bounds for every
$p\in[21/64,1]$, assuming only the classical edge bounds of planar graphs; the lower bound on paths,
and the upper bound for algorithms that know the length, in an explicit model of $r$-round
algorithms; the deterministic upper bound on forests and the size bound of the randomized one;
and the limits of the approach. Other methods reach ratio $7+\varepsilon$ in fewer rounds against the
integral optimum (arXiv:2610.31203) and ratio~$7$ in a constant number of rounds (arXiv:2611.02057);
we compare them in the introduction.
\end{abstract}

\section{Introduction}

\subsection{Dominating sets and the LOCAL model}

\begin{figure}[t]\centering
\begin{tikzpicture}
\begin{scope}[scale=1.0]
  \foreach \i/\x/\y in \exvertices \coordinate (a\i) at (\x,\y);
  \foreach \u/\w/\c in \exedges \draw[e] (a\u)--(a\w);
  \foreach \i in {1,3,4,5,6,8} \node[v] at (a\i) {};
  \foreach \i in {2,7} \node[d] at (a\i) {};
  \node[fcap] at (2,-0.85) {(a) a dominating set $D$ (black),\\ $\gamma=2$};
\end{scope}
\begin{scope}[shift={(6.6,1.6)}]
  \foreach \k in {0,...,4} \coordinate (c\k) at (90+72*\k:1.2);
  \draw[e] (c0)--(c1)--(c2)--(c3)--(c4)--cycle;
  \foreach \k in {0,...,4} \node[v,label={[lab]90+72*\k:$\frac13$}] at (c\k) {};
  \node[fcap] at (0,-2.45) {(b) $x_v=\frac13$ is optimal:\\ $\gamma_f(C_5)=\frac53<2=\gamma(C_5)$};
\end{scope}
\begin{scope}[shift={(11.6,1.6)},x=0.6cm,y=0.6cm]
  \fill[blue!8]  (0:2.45) \foreach \t in {60,120,...,300} {-- (\t:2.45)} -- cycle;
  \fill[blue!25] (0:1.45) \foreach \t in {60,120,...,300} {-- (\t:1.45)} -- cycle;
  \foreach \a in {-3,...,3} \foreach \b in {-3,...,3} {
    \pgfmathtruncatemacro\dd{max(abs(\a),abs(\b),abs(\a+\b))}
    \ifnum\dd<4
      \coordinate (t\a/\b) at ({\a+\b/2},{\b*0.866});
      \pgfmathtruncatemacro\ar{\a+1}\pgfmathtruncatemacro\ad{max(abs(\ar),abs(\b),abs(\ar+\b))}
      \ifnum\ad<4 \draw[faint] ({\a+\b/2},{\b*0.866})--({\ar+\b/2},{\b*0.866}); \fi
      \pgfmathtruncatemacro\bu{\b+1}\pgfmathtruncatemacro\bd{max(abs(\a),abs(\bu),abs(\a+\bu))}
      \ifnum\bd<4 \draw[faint] ({\a+\b/2},{\b*0.866})--({\a+\bu/2},{\bu*0.866}); \fi
      \pgfmathtruncatemacro\al{\a-1}\pgfmathtruncatemacro\cd{max(abs(\al),abs(\bu),abs(\al+\bu))}
      \ifnum\cd<4 \draw[faint] ({\a+\b/2},{\b*0.866})--({\al+\bu/2},{\bu*0.866}); \fi
    \fi}
  \foreach \a in {-3,...,3} \foreach \b in {-3,...,3} {
    \pgfmathtruncatemacro\dd{max(abs(\a),abs(\b),abs(\a+\b))}
    \ifnum\dd<4 \fill ({\a+\b/2},{\b*0.866}) circle (1.3pt); \fi}
  \node[d,minimum size=6pt,label={[lab,fill=blue!25,inner sep=1pt,label distance=1pt]below:$v$}] at (0,0) {};
  \node[fcap] at (0,-4.1) {(c) what $v$ knows after\\ $T=1$ and $T=2$ rounds};
\end{scope}
\end{tikzpicture}
\caption{Dominating sets and local views. In~(c), the shaded hexagons are the radius-$1$ and
radius-$2$ neighbourhoods of~$v$, from which a $1$-round and a $2$-round LOCAL algorithm decide
the output of~$v$.}
\label{fig:basics}
\end{figure}
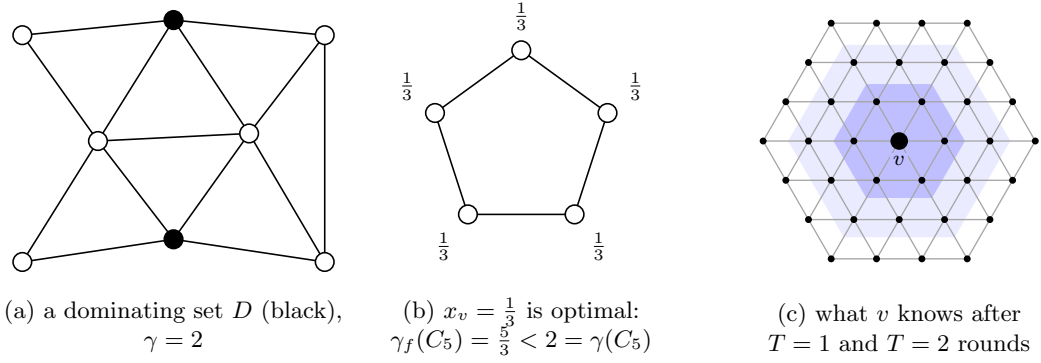

A set $D$ of vertices of a graph $G$ is \emph{dominating} if every vertex lies in $D$ or has a
neighbour in~$D$. The \emph{domination number} $\gamma(G)$ is the minimum size of a dominating set,
and minimum dominating set (MDS) is the problem of finding one. The integer program
$\min\sum_vx_v$ subject to $\sum_{u\in N[v]}x_u\ge1$ for every $v$, where $N[v]$ is the closed
neighbourhood of~$v$ and $x_v\in\{0,1\}$, has optimum $\gamma(G)$. Relaxing it to $x_v\ge0$ gives
the \emph{fractional domination number} $\gamma_f(G)\le\gamma(G)$ (Figure~\ref{fig:basics}).

MDS is NP-hard, even on planar graphs of maximum degree~$3$~\cite{GJ79}. It is a set cover problem
whose sets are the closed neighbourhoods. The greedy algorithm, which repeatedly adds a vertex that
dominates the most undominated vertices, therefore achieves ratio $1+\ln(\Delta+1)$ on graphs of
maximum degree~$\Delta$~\cite{J74}, and by the standard reduction from set cover, ratio
$(1-\varepsilon)\ln n$ is NP-hard~\cite{DS14}. Planar graphs are much easier: the layering technique
of~\cite{B94} gives a polynomial-time $(1+\varepsilon)$-approximation for every fixed $\varepsilon>0$.

In distributed computing the input graph is the network itself. Every vertex is a processor that
initially knows only its identifier and its incident edges, and it decides whether it belongs to $D$ by
exchanging messages with its neighbours. Dominating sets arise in this form as cluster heads in
wireless ad hoc and sensor networks, where every node needs a coordinator within one hop and no
node knows the whole topology. In the LOCAL model~\cite{L92}, computation proceeds in synchronous
rounds. Vertices have distinct identifiers from a finite range, and in a randomized algorithm every
vertex has its own infinite sequence of independent fair coins. In every round each vertex sends a message of arbitrary size to each neighbour and performs
arbitrary local computation. After $T$ rounds a vertex knows exactly its radius-$T$ neighbourhood,
with the identifiers and the coins in it, so a $T$-round algorithm is a function from radius-$T$
neighbourhoods to outputs, and the number of
rounds measures how far a vertex has to look. A constant-round algorithm is local in the strongest
sense. Its running time does not depend on the size of the network, a change in the graph affects
only the outputs within constant distance, and on graphs of bounded degree it yields a sublinear-time
algorithm that decides the output of a single vertex by exploring its neighbourhood~\cite{PR07}.

\subsection{Standard distributed algorithms}

\paragraph{Parallel greedy.} The greedy algorithm adds one vertex at a time. Its direct distributed
version lets a vertex join when its \emph{span}, the number of undominated vertices in its closed
neighbourhood, is the largest within distance~$2$, so that vertices joining in the same round do not
compete for the same undominated vertices. This can still take $\Omega(n)$ rounds, for example on a
caterpillar whose spine vertices have decreasing spans, because then only one vertex joins per
round. The algorithm of~\cite{JRS02} removes this bottleneck by rounding spans up to powers of~$2$,
so that many vertices tie. In every round, the vertices whose rounded span is maximal within
distance~$2$ become candidates, every undominated vertex counts the candidates that dominate it, and
every candidate $v$ joins with probability $1/\mathrm{med}(v)$, where $\mathrm{med}(v)$ is the
median of these counts over the undominated vertices that $v$ dominates. The expected size of the
result is $O(\log\Delta)\gamma(G)$, and the algorithm terminates in $O(\log n\log\Delta)$ rounds
with high probability.

\paragraph{LP relaxation and rounding.} A fractional dominating set $x$ becomes a dominating set in
two rounds. Every vertex joins independently with probability $\min\{1,x_v\ln(\Delta+1)\}$, and
every vertex that is still undominated joins as well. Since $\sum_{u\in N[v]}x_u\ge1$, a vertex
remains undominated after the first step with probability at most $e^{-\ln(\Delta+1)}=1/(\Delta+1)$.
Every vertex dominates at most $\Delta+1$ vertices, so $n/(\Delta+1)\le\gamma_f(G)$, and the expected
size is at most $\ln(\Delta+1)\sum_vx_v+\gamma_f(G)$. In~\cite{KMW16}, covering and packing LPs are
solved within a factor $1+\varepsilon$ in $O(\log n/\varepsilon)$ rounds. Together with a rounding of
this kind, this gives an $O(\log\Delta)$-approximation of MDS in $O(\log n)$ rounds. In the same
paper it was shown that every polylogarithmic approximation, even a randomized one, needs
$\Omega\big(\min\{\log\Delta/\log\log\Delta,\sqrt{\log n/\log\log n}\}\big)$ rounds. So on general
graphs no constant-round algorithm has a constant ratio.

\paragraph{Planar graphs in a constant number of rounds.} Sparse classes behave differently. The
algorithm of~\cite{LOW08} has two steps. First, every vertex $v$ whose open neighbourhood cannot be
dominated by six vertices other than~$v$ joins~$D$. This needs only the radius-$2$ neighbourhood
of~$v$. Second, every vertex that is still undominated adds a vertex of its closed neighbourhood that
dominates the most undominated vertices. The algorithm runs in six rounds and has ratio $130$ on
planar graphs~\cite{LOW08}. Planarity enters only in the analysis, which shows that the first step
selects fewer than $3\gamma(G)$ vertices outside a minimum dominating set. This raised the question of the best ratio that a constant
number of rounds can achieve on planar graphs (Section~\ref{sec:optimal}).

\begin{figure}[t]\centering
\begin{tikzpicture}[scale=0.8]
\draw[fill=blue!12,draw=blue!50] (0,0) circle (1.1);
\foreach \t in {120,240} {
  \draw[fill=orange!15,draw=orange!60] (\t:2.5) circle (1.1);
  \fill[black!45] (\t:2.5) circle (2pt);
  \draw[e] (\t:1.1)--(\t:1.4);}
\draw[fill=orange!15,draw=orange!60] (3.1,0) circle (1.1);
\fill[black!45] (3.1,0) circle (2pt);
\node[d,label={[lab]left:leader}] (l) at (0,0) {};
\coordinate (x) at (1.1,0); \coordinate (y) at (2.0,0); \coordinate (z) at (4.2,0);
\draw[e] (l)--(x); \draw[line width=1.2pt] (x)--(y); \draw[e] (y)--(z);
\foreach \p in {x,y,z} \fill (\p) circle (1.6pt);
\foreach \p in {0,1.1,2.0,4.2} \draw[line width=0.4pt] (\p,0.22)--(\p,0.48);
\foreach \a/\b/\t in {0/1.1/{$\le r$},1.1/2.0/{$1$},2.0/4.2/{$\le 2r$}}
  \draw[<->,lab,shorten <=1.2pt,shorten >=1.2pt] (\a,0.35)--node[above]{\t} (\b,0.35);
\draw[|<->|,lab] (0,-1.45)--node[below]{$\le 3r+1$} (4.2,-1.45);
\node[lab,blue!60!black] at (-0.15,-0.62) {centre part};
\node[lab,orange!60!black] at (3.1,-0.72) {leaf part};
\end{tikzpicture}
\caption{A star of parts of radius at most $r$ becomes one part of radius at most $3r+1$ around the
leader of its centre part.}
\label{fig:radius}
\end{figure}
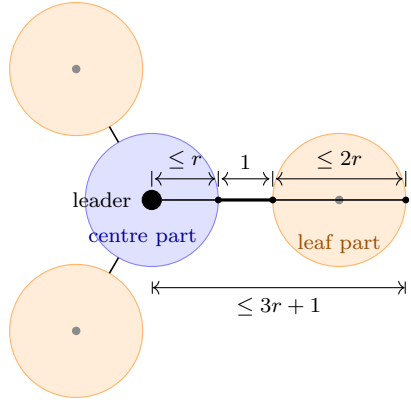

\paragraph{Cluster-and-solve.} Ratios close to the optimum need a global view of larger regions.
Algorithms of this kind on sparse classes~\cite{CHW08,LR15} partition the graph into clusters of
small radius such that few edges run between clusters, let one vertex of every cluster collect the
cluster and solve it exactly, and combine the local solutions. The quality of the result depends on
the weight of the edges between clusters, and the number of rounds on the cluster radius.
In~\cite{CHW08}, this gives a $(1+\varepsilon)$-approximation of MDS on planar graphs in $O(\log^*n)$
rounds. A common way to build the clusters is repeated star contraction: every cluster picks a
neighbouring cluster to merge with, such that the merges form vertex-disjoint stars. Every star
becomes one cluster, and parts of radius at most $r$ give a cluster of radius at most $3r+1$
(Figure~\ref{fig:radius}).

\subsection{Optimal ratio on planar graphs}\label{sec:optimal}

In~\cite{HLS14} it was shown that no deterministic constant-round algorithm achieves ratio
$7-\varepsilon$ on planar graphs. A sequence of works
improved the upper bound~\cite{HKOSV22,W26}, and the gap was recently closed in~\cite{AS26}: for
every $\varepsilon>0$ there is a deterministic algorithm that outputs a dominating set of size at most $(7+\varepsilon)\gamma_f(G)$ in
\[
  T_0+O\big((c_0/\varepsilon)^{\kappa}\big)\ \text{rounds},\qquad
  \kappa=\frac{\ln 3}{\ln(1/q)},\quad q=1-\frac{1}{8a}=\frac{23}{24},
\]
where $a=3$ is the edge density of planar graphs (every subgraph $H$ has at most $a|V(H)|$ edges) and $T_0,c_0$ are absolute constants. The exponent
$\kappa\approx 25.81$ comes from a randomized star-contraction clustering~\cite{LR15,CHW08}:
each level grows part radii by a factor of~$3$ and reduces the expected weight between parts by a factor~$q$.
Bringing this weight down to an $\varepsilon$-fraction takes $L\approx\ln(1/\varepsilon)/\ln(1/q)$
levels, and hence radius and round number $3^L\approx(1/\varepsilon)^{\kappa}$.
Open Problem~4 of~\cite{AS26} asks whether this exponent can be substantially reduced.

\paragraph{Our result.} We change only the star-selection rule of the clustering step
(Section~\ref{sec:rule}). For planar graphs and $p\in[21/64,1]$, the per-level reduction factor
improves from $23/24$ to $1-\varphi(p)$ with $\varphi(p)=(1-p)\big(3-2(1-p)^3-(1-p)^6\big)/9$
(Theorem~\ref{thm:planar}). For a class of density $a$, it improves from $1-1/(8a)$ to
$1-f_k(p)$ with $f_k(p)=(1-p)(1-(1-p)^k)/k$ and $k=\lceil a\rceil$ (Theorem~\ref{thm:level}).
Everything else in~\cite{AS26} carries over unchanged (Section~\ref{sec:integration}):
the radius recursion, the view radius, the derandomization and the rounding.

\begin{theorem}\label{thm:main}
For every $\varepsilon\in(0,1]$ there is a deterministic LOCAL algorithm that returns, on every
planar graph $G$, a dominating set of size at most $(7+\varepsilon)\gamma_f(G)$ in
$T_0+O((c_0/\varepsilon)^{\kappa'})$ rounds, where $\kappa'=\ln 3/\ln(2097152/1740037)<5.89$, and
$T_0,c_0$ are as in~\cite[Cor.~8.1]{AS26}. With one fair coin per part and level, as
in~\cite{AS26}, the same holds with $\kappa'=\ln3/\ln(1152/977)<6.7$, and with six fair coins
($p=21/64$) it holds with $\kappa'<5.83$.
\end{theorem}

\begin{figure}[t]\centering
\begin{tikzpicture}[x=2.35cm]
\draw[line width=0.8pt] (0,0)--(5.2,0);
\draw[line width=0.8pt] (5.55,0)--(6.05,0);
\foreach \s in {5.25,5.4} \draw (\s,-0.12)--(\s+0.08,0.12);
\foreach \k in {2,...,7} {\draw (\k-2,0)--(\k-2,-0.08); \node[font=\tiny,below] at (\k-2,-0.08) {\k};}
\draw (5.8,0)--(5.8,-0.08); \node[font=\tiny,below] at (5.8,-0.08) {25.81};
\node[lab,right] at (6.05,0) {$\kappa$};
\tikzset{ub/.style={blue!65!black},lb/.style={red!65!black},
  t/.style={lab,align=center}}
\foreach \k/\h/\txt in {3.89/0.55/{this work, $p=\frac38$\\ $\kappa<5.89$},
                        4.70/1.25/{this work, $p=\frac12$\\ $\kappa<6.7$},
                        5.8/0.55/{original\\ $\kappa\approx25.81$}} {
  \draw[ub,line width=0.7pt] (\k,0)--(\k,\h);
  \fill[ub] (\k,0) circle (1.7pt);
  \node[t,ub,above] at (\k,\h) {\txt};}
\tikzset{lnw/.style={anchor=north west,align=left},lne/.style={anchor=north east,align=right},ln/.style={anchor=north,align=center}}
\foreach \k/\h/\o/\txt in {3.74/0.75/lnw/{random-centre rule,\\ join edges only},
                           3.14/1.75/lnw/{random-centre rule,\\ all edges},
                           2.74/1.75/lne/{independent coins,\\ join edges only},
                           1.82/0.75/ln/{any star contraction,\\ join edges only},
                           0.15/0.75/ln/{any star contraction,\\ per-level worst case}} {
  \draw[lb,line width=0.7pt] (\k,0)--(\k,-\h);
  \draw[lb,fill=white,line width=0.7pt] (\k,0) circle (1.7pt);
  \node[lab,lb,inner sep=2pt,\o] at (\k,-\h) {\txt};}
\end{tikzpicture}
\caption{Round exponents for planar graphs. Above the axis: proved upper bounds on~$\kappa$. Below
the axis: the smallest exponent that the indicated kind of analysis can give
(Section~\ref{sec:limits}).}
\label{fig:exponents}
\end{figure}
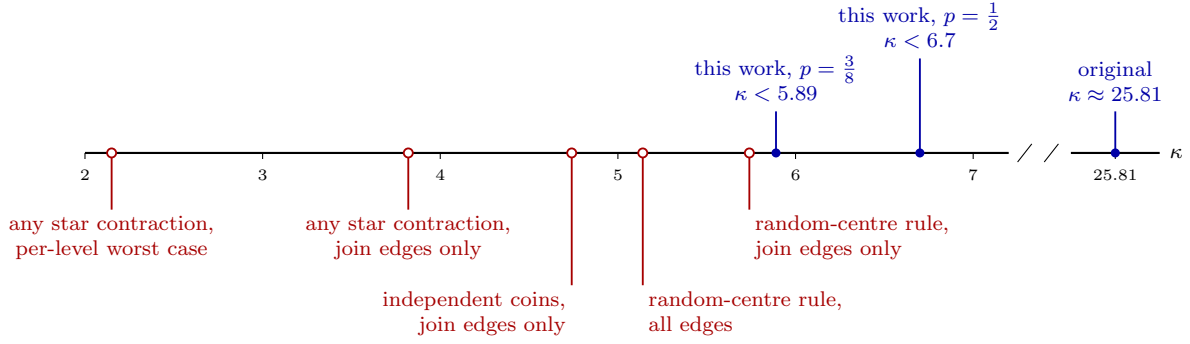

The first bound uses three fair coins per part and level ($p=3/8$). The six-coin exponent is within
$10^{-3}$ of the best exponent that Theorem~\ref{thm:planar} gives for any~$p$.
Figure~\ref{fig:exponents} compares the exponents for $p=3/8$ and $p=1/2$ with the limits of
Section~\ref{sec:limits}.

For a general proper minor-closed class of density $a$, the exponent $\ln 3/\ln(1/q)$ improves
as in Table~\ref{tab:general}. Lemma~\ref{lem:dominates} shows that the new factor is never worse.

\begin{table}[ht]\centering
\begin{tabular}{rrrrr}\toprule
$a$ & \cite{AS26}: $q=1-\frac1{8a}$ & ours, $p=\frac12$ & ours, best dyadic $p$ & improvement\\\midrule
1 & 8.23 & 3.82 & 3.82 & $2.15\times$\\
2 & 17.02 & 5.30 & 5.14 & $3.31\times$\\
3 (planar) & 25.81 & 6.97 & 6.42 & $4.02\times$\\
6 & 52.18 & 12.84 & 10.08 & $5.17\times$\\
10 & 87.34 & 21.44 & 14.81 & $5.89\times$\\
100 & 878.34 & 219.18 & 115.67 & $7.59\times$\\\bottomrule
\end{tabular}
\caption{Round exponent $\kappa=\ln3/\ln(1/q)$ for integer density $a$ (so $k=a$).
The best dyadic $p$ uses at most $8$ fair coins per part and level. Our exponents are rounded up,
the improvement factors down. For planar graphs,
Theorem~\ref{thm:planar} improves the row $a=3$ to $6.67$ ($p=1/2$) and $5.83$ ($p=21/64$).}
\label{tab:general}
\end{table}

\paragraph{Positioning.} The improvement concerns the optimal-ratio regime. The algorithm
of~\cite{W26} is a deterministic $(8+\varepsilon)$-approximation whose last phase is that
of~\cite{HKOSV22}. That phase reduces to maximum degree $\Gamma=O(1/\varepsilon)$ by selecting
vertices at a constant-factor cost, and then runs the LP algorithm
of~\cite{KMW16} in $O(r^2)$ rounds with $r=\log\Gamma/\log(1+\varepsilon)$, roughly
$(\log(1/\varepsilon)/\varepsilon)^2$. That degree reduction loses a constant factor. To reach ratio~$7$,
the algorithm of~\cite{AS26} instead computes a $(1+\varepsilon)$-approximation of $\gamma_f$ on the
whole graph by cluster-and-solve, and Theorem~\ref{thm:main} is the fastest version of this algorithm
known to us. The same holds for the other guarantees of~\cite[Thm.~6.2]{AS26} that our substitution
improves for every proper minor-closed class: the deterministic $(1+\varepsilon)$-approximation of
$\gamma_f$, the randomized algorithm with expected size $(1+\varepsilon)\gamma$, and the ratio
$2a+1+\varepsilon$. On forests and on outerplanar graphs, Theorems~\ref{thm:forests}
and~\ref{thm:outerplanar} give much faster algorithms for the first two.

\paragraph{Other methods.} Two recent works reach ratio~$7$ on planar graphs by other methods, with
fewer rounds than Theorem~\ref{thm:main} for small~$\varepsilon$. The algorithm of~\cite{P26} is a deterministic
$(7+\varepsilon)$-approximation against the integral optimum~$\gamma$ in
$O(\varepsilon^{-4}\log^2(1/\varepsilon))$ rounds, and it extends beyond minor-closed classes. The
algorithm of~\cite{ASW26} outputs a dominating set of size at most $7\gamma_f$ in a constant number
of rounds, in the broadcast CONGEST model, and in general of size at most $(2a+1)\gamma_f$ for
rational~$a$. That work also gives a deterministic constant-round $(1+\varepsilon)$-approximation of
$\gamma_f$ and a randomized constant-round algorithm with expected size $(1+\varepsilon)\gamma$. As
we read it, their round bounds depend only on the fixed parameters and are not stated as explicit
functions of~$\varepsilon$. So Theorem~\ref{thm:main} and Table~\ref{tab:general} concern the
algorithm of~\cite{AS26}, and they give explicit bounds on the rounds of its
$(1+\varepsilon)$-approximations. The lower bound of Theorem~\ref{thm:lower} holds for every
algorithm of these two kinds, including those of~\cite{ASW26}. The optimal dependence on~$\varepsilon$
on planar graphs remains open (Section~\ref{sec:open}).

\paragraph{The dependence on $\varepsilon$.} Open Problem~3 of~\cite{AS26} asks for the optimal
dependence on $\varepsilon$ of the deterministic $(1+\varepsilon)$-approximation of $\gamma_f$ and of
the randomized algorithm with expected size $(1+\varepsilon)\gamma$. On paths, both need at least
$1/\varepsilon-4$ rounds (Theorem~\ref{thm:lower}), so $\Omega(1/\varepsilon)$ rounds are necessary
on every minor-closed class that contains all paths. On forests (Theorem~\ref{thm:forests}) and on
outerplanar graphs of arbitrary maximum degree (Theorem~\ref{thm:outerplanar}), both run in
$O(1/\varepsilon)$ rounds, so the answer there is $\Theta(1/\varepsilon)$. The algorithms average the
optimal solutions of overlapping windows of radius $O(1/\varepsilon)$. On outerplanar graphs the
windows live in a covering graph in which every long face unrolls into an infinite path.

\paragraph{Formal verification.} The bounds for one level of the rule (Section~\ref{sec:rule}), in the
planar case for every $p\in[21/64,1]$ and on the surfaces of Table~\ref{tab:surfaces}, the resulting
exponents and their limit as the density grows (Proposition~\ref{prop:eight}), the entries of
Table~\ref{tab:general}, Proposition~\ref{prop:kkn} and Theorem~\ref{thm:lower} are checked in Lean~4
with Mathlib (Section~\ref{sec:lean}). The direction
of the orientation theorem of~\cite{H65} that we use and Lemmas~\ref{lem:facts}
and~\ref{lem:factsgen} are checked as well. Hence the level bounds assume only the classical edge
bounds of planar graphs from Euler's formula: at most $3n$ edges, and at most $2n$ if the graph is
bipartite. Theorem~\ref{thm:lower} is checked in a model of
$r$-round algorithms on paths, for deterministic and for randomized algorithms, and so is the
upper bound $1+2/(2r+3)$ on paths for algorithms that know the length. The integration into the algorithm of~\cite{AS26} is proved on paper, except
for the step from the per-level factor to the exponent. The deterministic part of
Theorem~\ref{thm:forests} is checked as well, with its explicit constants. For its randomized part,
Lean checks the expected size, the cut probability and the dependence on the random values, with the
input of~\cite{LOW08} as a hypothesis; that the clusters are computed locally is proved on paper.
Theorem~\ref{thm:outerplanar} is proved on paper, and we check its construction numerically. Lean
checks the steps that do not depend on the graph class (the packing bound, the repair and the
averaging over windows), the run lemmas, the interval facts of the randomized version and the
arithmetic of all constants, with the structure of the covering graph as hypotheses. The limits of
Section~\ref{sec:limits} are checked as well, with the planarity of the graphs used there taken as
given.

\section{One level of random-centre contraction}\label{sec:rule}

\begin{figure}[t]\centering
\begin{tikzpicture}
\begin{scope}[scale=0.85]
  \foreach \i/\x/\y in \exvertices \coordinate (a\i) at (\x,\y);
  \foreach \u/\w/\c in \exedges \draw[e] (a\u)--(a\w) node[midway,wl] {\c};
  \foreach \i in {1,3,4,5,6,8} \node[v] at (a\i) {};
  \foreach \i in {2,7} \node[d] at (a\i) {};
  \node[fcap] at (2,-0.85) {(a) parts whose coins come up\\ heads become centres (black)};
\end{scope}
\begin{scope}[shift={(4.9,0)},scale=0.85]
  \foreach \i/\x/\y in \exvertices \coordinate (b\i) at (\x,\y);
  \foreach \u/\w/\c in \exedges \draw[faint] (b\u)--(b\w) node[midway,wl,text=black!45] {\c};
  \foreach \u/\w/\c in {1/2/3,3/2/1,5/2/4,4/7/2,6/7/2,8/7/1} \draw[jn] (b\u)--(b\w) node[midway,wl] {\c};
  \foreach \i in {1,3,4,5,6,8} \node[v] at (b\i) {};
  \foreach \i in {2,7} \node[d] at (b\i) {};
  \node[fcap] at (2,-0.85) {(b) every other part joins its\\ heaviest centre neighbour};
\end{scope}
\begin{scope}[shift={(9.8,0)},scale=0.85]
  \foreach \i/\x/\y in \exvertices \coordinate (c\i) at (\x,\y);
  \begin{scope}[line width=17pt,line cap=round,line join=round]
    \draw[blue!15] (c1)--(c2)--(c5) (c2)--(c3);
    \draw[orange!22] (c4)--(c7)--(c8) (c7)--(c6);
  \end{scope}
  \foreach \u/\w in {1/2,2/3,2/5,3/5,4/6,4/7,6/7,7/8} \draw[e] (c\u)--(c\w);
  \foreach \u/\w/\c in {1/4/2,2/4/1,4/5/1,5/7/1,5/8/3,3/8/2} \draw[cut] (c\u)--(c\w) node[midway,wl] {\c};
  \foreach \i in {1,3,4,5,6,8} \node[v] at (c\i) {};
  \foreach \i in {2,7} \node[d] at (c\i) {};
  \node[fcap] at (2,-0.85) {(c) two new parts; the weight\\ between parts drops from 28 to 10};
\end{scope}
\end{tikzpicture}
\caption{One level of the random-centre rule on a quotient $Q$ of total weight~$28$. The join edges
(bold in~(b)) carry weight~$13$. The two edges between leaves of the same star, of weights $2$
and~$3$, become internal as well.}
\label{fig:level}
\end{figure}
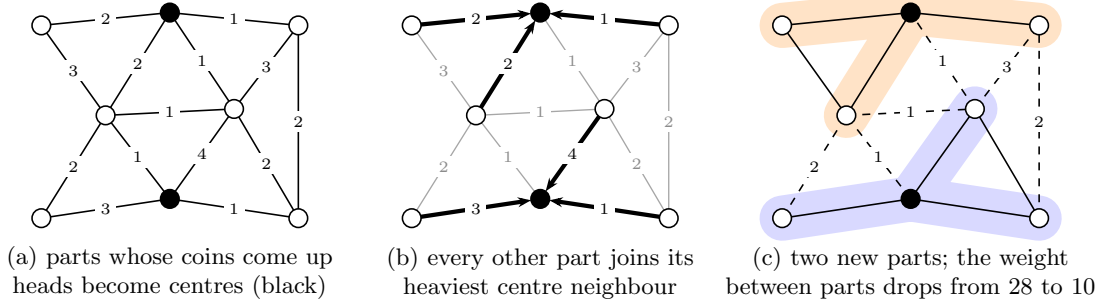

Let $Q$ be a finite simple graph with edge weights $w\colon E(Q)\to\mathbb{R}_{\ge0}$ and distinct
identifiers. For $C\subseteq V(Q)$, a vertex $P\notin C$ with a neighbour in $C$ \emph{joins} a
neighbour $c(P)\in C$ that maximizes $w(Pc(P))$, breaking ties by identifier. The \emph{join edges}
are the edges $Pc(P)$, and the \emph{gain} is $\mathrm{gain}(C)=\sum_{P\notin C}\max_{X\in C\cap
N(P)}w(PX)$, with $\max\emptyset=0$. Figure~\ref{fig:level} shows an example.

\begin{lemma}\label{lem:distinct}
Distinct joining vertices have distinct join edges, so $\mathrm{gain}(C)$ is the total weight of a
set of edges. Contracting each centre together with the vertices that joined it yields
vertex-disjoint stars, and every join edge becomes internal.
\end{lemma}
\begin{proof}
A join edge $Pc(P)$ has exactly one endpoint outside $C$, namely $P$. Hence $Pc(P)=P'c(P')$ forces
$P=P'$. Centres do not join, and every non-centre joins at most one centre.
\end{proof}

For a finite set $A$ let $C_A\subseteq A$ contain each element independently with probability $p$.
Write $s=1-p$ and $g(m)=1-s^m$.

\begin{lemma}[vertex lemma]\label{lem:vertex}
For $p\in[0,1]$ and weights $w\colon A\to\mathbb{R}_{\ge0}$,
\[
  |A|\cdot\E\Big[\max_{x\in C_A}w(x)\Big]\ \ge\ g(|A|)\sum_{x\in A}w(x).
\]
\end{lemma}
\begin{proof}
We use induction on $n+1=|A|$; the case $A=\emptyset$ is trivial. Let $y\in A$ have maximum weight,
and put $A'=A\setminus\{y\}$, $W'=\sum_{x\in A'}w(x)$ and
$E'=\E[\max_{x\in C_{A'}}w(x)]$. Conditioning on whether $y\in C_A$ gives
$\E[\max_{C_A}w]=sE'+pw(y)$, because $y$ is the heaviest centre whenever it is a centre. The
induction hypothesis gives $nE'\ge g(n)W'$, and the choice of $y$ gives $W'\le nw(y)$. Using
$g(n+1)=1-s\,s^n$, a direct expansion verifies
\[
  n\Big[(n+1)\big(sE'+pw(y)\big)-g(n+1)\big(w(y)+W'\big)\Big]
  =(n+1)s\big(nE'-g(n)W'\big)+\big(nw(y)-W'\big)\big(np-s\,g(n)\big).
\]
Both terms are non-negative. For the second, Bernoulli's inequality gives $g(n)\le np$, and
$s\le 1$. For $n\ge1$ we divide by $n$. For $n=0$ we have $W'=0$, and the claim reads
$pw(y)\le sE'+pw(y)$.
\end{proof}

Since $g$ is concave with $g(0)=0$, the ratio $g(m)/m$ is non-increasing. Hence $g(m)/m\ge g(k)/k$ for $1\le m\le k$.

\begin{theorem}\label{thm:level}
Suppose the edges of $Q$ admit an orientation in which every vertex has out-degree at most $k\ge1$.
If every vertex of $Q$ is in $C$ independently with probability $p\in[0,1]$, then
\[
  \E[\mathrm{gain}(C)]\ \ge\ f_k(p)\,w(Q),\qquad f_k(p)=\frac{(1-p)\big(1-(1-p)^k\big)}{k}.
\]
\end{theorem}
\begin{proof}
Let $O(P)$ be the out-neighbourhood of $P$, so $w(Q)=\sum_P\sum_{X\in O(P)}w(PX)$. Fix $P$. Then
$\max_{X\in C\cap N(P)}w(PX)\ge\max_{X\in C\cap O(P)}w(PX)$. The event $P\notin C$ is independent
of $C\cap O(P)$, because $P\notin O(P)$. If $O(P)=\emptyset$, the sum below is empty. Otherwise,
by Lemma~\ref{lem:vertex} applied to $A=O(P)$ and by the monotonicity of $g(m)/m$,
\[
  \E\Big[[P\notin C]\max_{X\in C\cap N(P)}w(PX)\Big]\ \ge\ (1-p)\frac{g(|O(P)|)}{|O(P)|}\sum_{X\in O(P)}w(PX)
  \ \ge\ f_k(p)\sum_{X\in O(P)}w(PX).
\]
Summing over $P$ proves the claim.
\end{proof}

By~\cite{H65}, a graph has an orientation with out-degree at most $k$ if and only if every
subgraph $H$ satisfies $|E(H)|\le k|V(H)|$. A graph in a class of density $a$ therefore admits such
an orientation with $k=\lceil a\rceil$. For planar graphs $k=3$. Over $p\in[0,1]$, $f_3$ is maximized at
$p=1-4^{-1/3}\approx0.370$. The dyadic choice $p=3/8$ (three fair coins) gives
$f_3(3/8)=1935/12288\approx0.15747$, and $f_3(p)\le 63/400$ for every~$p$, so the loss is below $3\cdot10^{-5}$.

\begin{lemma}\label{lem:dominates}
For every $a\ge1/2$, $f_{\lceil a\rceil}(1/2)\ge 1/(8a)$.
\end{lemma}
\begin{proof}
Let $k=\lceil a\rceil\ge1$. Then $f_k(1/2)=(1-2^{-k})/(2k)\ge 1/(4k)$, and $k\le 2a$: if $k=1$ this
follows from $a\ge1/2$, and if $k\ge2$ from $k<a+1\le 2a$.
\end{proof}

As the density grows, the improvement approaches a factor of~$8$, provided that $p$ shrinks at the
right speed.

\begin{proposition}\label{prop:eight}
Let $\kappa_{\mathrm{AS}}(a)=\ln3/\ln\big(1/(1-\frac1{8a})\big)$ and
$\kappa(a,p)=\ln3/\ln\big(1/(1-f_{\lceil a\rceil}(p))\big)$. If $p=p(a)\in[0,1]$ satisfies $p\to0$
and $ap\to\infty$ as $a\to\infty$, then $\kappa_{\mathrm{AS}}(a)/\kappa(a,p(a))\to8$. This holds for
$p=2^{-j}$ with $j=\lfloor\frac12\lfloor\log_2\lceil a\rceil\rfloor\rfloor$, that is, with $j$ fair
coins per part and level. For a fixed $p\in(0,1]$, the ratio tends to $8(1-p)$.
\end{proposition}
\begin{proof}
Let $k=\lceil a\rceil$, $y=1/(8a)$ and $f=f_k(p)$. For $0\le x<1$ we have
$x\le\ln(1/(1-x))\le x/(1-x)$, so the ratio $\ln(1/(1-f))/\ln(1/(1-y))$ lies between $8af(1-y)$ and
$8af/(1-f)$. Here $8af=8\frac ak(1-p)\big(1-(1-p)^k\big)$, $a/k\to1$, and
$(1-p)^k\le e^{-pk}\le e^{-ap}\to0$. Hence $8af\to8$, and $f=8af\cdot y\to0$, so both bounds tend
to~$8$. For a fixed $p$, $(1-p)^k\to0$ because $k\to\infty$, and $8af\to8(1-p)$. For the dyadic
choice, $4^j\le k<4^{j+1}$ gives $p^2<4/a$ and $(ap)^2\ge a^2/(a+1)\ge a/2$ for $a\ge1$.
\end{proof}

\subsection{A sharper count for planar graphs}\label{sec:planar}

Theorem~\ref{thm:level} charges every edge to one endpoint only. For planar quotients we count every
edge at both endpoints. For $\tau>0$ let $Q_\tau$ be the spanning subgraph of the edges of weight at
least~$\tau$, and let $d_\tau(P)$ be the degree of $P$ in~$Q_\tau$.

\begin{lemma}[layer cake]\label{lem:cake}
If every vertex of $Q$ is in $C$ independently with probability $p$, then
\[
  \E[\mathrm{gain}(C)]=s\int_0^\infty\sum_P g(d_\tau(P))\,d\tau.
\]
\end{lemma}
\begin{proof}
For fixed $P$, $[P\notin C]\max_{X\in C\cap N(P)}w(PX)=\int_0^\infty[P\notin C]\,[C\cap N_{Q_\tau}(P)\ne\emptyset]\,d\tau$.
The event $P\notin C$ has probability $s$ and is independent of $C\cap N(P)$, and
$\Pr[C\cap N_{Q_\tau}(P)\ne\emptyset]=g(d_\tau(P))$. Take expectations, exchange them with the
integral (the integrand is bounded and vanishes for $\tau>\max w$), and sum over~$P$.
\end{proof}

\begin{lemma}\label{lem:facts}
Every simple planar graph with degrees $d_v$ and $n'$ non-isolated vertices satisfies
\[
  \text{(A)}\ \ |E|\le\sum_v\min(d_v,3),\qquad
  \text{(B)}\ \ 2|E|\le\sum_v\min(d_v,6)+2n'.
\]
\end{lemma}
\begin{proof}
(A) Let $H=\{v:d_v\ge3\}$. At most $3|H|$ edges lie inside $H$, because every simple planar graph on
$m$ vertices has at most $3m$ edges. Every other edge has an endpoint $v\notin H$, and there are at
most $\sum_{v\notin H}d_v=\sum_{v\notin H}\min(d_v,3)$ of them.
(B) Let $H=\{v:d_v\ge6\}$. Then $\sum_{v\in H}d_v=2|E(H)|+|E(H,\bar H)|\le 6|H|+|E(H,\bar H)|$, so
$2|E|=\sum_v\min(d_v,6)+\sum_{v\in H}(d_v-6)\le\sum_v\min(d_v,6)+|E(H,\bar H)|$. The edges between $H$
and $\bar H$ form a simple bipartite planar graph on non-isolated vertices, which has at most
$2n'$ edges.
\end{proof}

\begin{lemma}[degree inequality]\label{lem:degree}
Let $p\in[21/64,1]$ and $c(s)=\big(2g(3)+g(6)\big)/9=(3-2s^3-s^6)/9$. Every simple planar graph
satisfies $\sum_v g(d_v)\ge c(s)\,|E|$.
\end{lemma}
\begin{proof}
Put $\beta=(g(6)-g(3))/3$ and $\alpha=(8g(3)-5g(6))/9$, so that $\alpha+2\beta=c(s)$. Here
$\beta\ge0$ because $g$ is increasing, and $\alpha\ge0$ because $9\alpha=(1-s^3)(3-5s^3)$ and
$s^3\le 3/5$. We claim that for every
integer $d\ge0$,
\[
  g(d)\ \ge\ \alpha\min(d,3)+\beta\big(\min(d,6)+2[d\ge1]\big).
\]
Equality holds at $d\in\{0,3,6\}$. For $d>6$ the right side is constant and $g$ is increasing. For
$d\in\{4,5\}$ the right side is the chord of the concave function $g$ between $3$ and $6$. For
$d\in\{1,2\}$ the claim reads $9g(1)\ge4g(6)-g(3)$ and $9g(2)\ge4g(3)+2g(6)$, that is,
$h_1(s)=6-9s-s^3+4s^6\ge0$ and $h_2(s)=3-9s^2+4s^3+2s^6\ge0$. Both polynomials are decreasing on
$[0,43/64]$ ($h_1'\le-9+24s^5<0$, and $h_2'=6s(2s+2s^4-3)<0$), and $h_1(43/64)>0.017$,
$h_2(43/64)>0.33$. Summing the claim over all vertices and applying Lemma~\ref{lem:facts} gives
$\sum_vg(d_v)\ge\alpha|E|+2\beta|E|$.
\end{proof}

\begin{theorem}\label{thm:planar}
Let $Q$ be a simple planar graph. If every vertex of $Q$ is in $C$ independently with probability
$p\in[21/64,1]$, then
\[
  \E[\mathrm{gain}(C)]\ \ge\ \varphi(p)\,w(Q),\qquad \varphi(p)=\frac{(1-p)\big(3-2(1-p)^3-(1-p)^6\big)}{9}.
\]
\end{theorem}
\begin{proof}
Every $Q_\tau$ is a simple planar graph, and $\int_0^\infty|E(Q_\tau)|\,d\tau=w(Q)$. Apply
Lemma~\ref{lem:degree} to each $Q_\tau$ inside Lemma~\ref{lem:cake}.
\end{proof}

The values are $\varphi(3/8)=357115/2097152\approx0.17029$, $\varphi(1/2)=175/1152\approx0.15191$
and $\varphi(21/64)\approx0.17181$, compared with $f_3(3/8)\approx0.15747$ and $f_3(1/2)=7/48$.
Over $p\in[0,1]$, $\varphi$ is maximized at $s^3=(\sqrt{37}-4)/7$, that is, $p\approx0.3324$, where
$\varphi\approx0.17182$. The count is tight for its inputs. Over degree sequences that satisfy only (A) and (B), the infimum
of $\sum_vg(d_v)/|E|$ is $c(s)$. It is approached by two vertices of degree~$3$ per vertex of
degree~$6$, together with a vanishing number of hubs incident to two thirds of the edges. The
certificate above is the dual solution of this linear program.

\subsection{Other classes}\label{sec:classes}

Planarity enters Theorem~\ref{thm:planar} only through Lemma~\ref{lem:facts}, and its proof uses
two density bounds. Let $\mathcal C$ be a class of graphs that is closed under taking subgraphs,
and let $a,b>0$ be such that every $H\in\mathcal C$ has $|E(H)|\le a|V(H)|$ and every bipartite
$H\in\mathcal C$ has $|E(H)|\le b|V(H)|$. Planar graphs have $a=3$ and $b=2$.

\begin{lemma}\label{lem:factsgen}
Every $H\in\mathcal C$ with degrees $d_v$ and $n'$ non-isolated vertices satisfies, for all reals
$t\ge a$ and $u\ge 2a$,
\[
  \text{(A$_t$)}\ \ |E|\le\sum_v\min(d_v,t),\qquad
  \text{(B$_u$)}\ \ 2|E|\le\sum_v\min(d_v,u)+b\,n'.
\]
\end{lemma}
\begin{proof}
As for Lemma~\ref{lem:facts}. At most $a|H_t|\le\sum_{v\in H_t}\min(d_v,t)$ edges lie inside
$H_t=\{v:d_v\ge t\}$. For $H_u=\{v:d_v\ge u\}$ we have
$\sum_{v\in H_u}d_v\le 2a|H_u|+|E(H_u,\bar H_u)|\le u|H_u|+|E(H_u,\bar H_u)|$, and the edges between
$H_u$ and $\bar H_u$ form a bipartite graph in $\mathcal C$ on at most $n'$ non-isolated vertices.
\end{proof}

\begin{theorem}\label{thm:general}
Let $s=1-p$. Let $\alpha_1,\dots,\alpha_m,\beta_1,\dots,\beta_\ell\ge0$ and $t_i\ge a$,
$u_j\ge2a$ satisfy, for every integer $d\ge0$,
\[
  g(d)\ \ge\ \sum_i\alpha_i\min(d,t_i)+\sum_j\beta_j\big(\min(d,u_j)+b\,[d\ge1]\big).
\]
If every vertex of a simple graph $Q\in\mathcal C$ is in $C$ independently with probability $p$,
then $\E[\mathrm{gain}(C)]\ge s\,c\,w(Q)$ with $c=\sum_i\alpha_i+2\sum_j\beta_j$.
\end{theorem}
\begin{proof}
Every $Q_\tau$ lies in $\mathcal C$. Summing the certificate over its vertices and applying
Lemma~\ref{lem:factsgen} gives $\sum_Pg(d_\tau(P))\ge c\,|E(Q_\tau)|$. Integrate as in
Theorem~\ref{thm:planar}.
\end{proof}

Lemma~\ref{lem:degree} is the case $t=3$, $u=6$, $b=2$. Without any bipartite bound, a single
threshold already improves Theorem~\ref{thm:level} when $a$ is not an integer.

\begin{corollary}\label{cor:aonly}
Let $a\ge1$ and $c_a(s)=\min\{g(\lfloor a\rfloor)/\lfloor a\rfloor,\ g(\lceil a\rceil)/a\}$. Then
$\E[\mathrm{gain}(C)]\ge s\,c_a(s)\,w(Q)$ for every simple $Q\in\mathcal C$. For integer $a$ this is
$f_a(p)$. For non-integer $a$ and $0<s<1$ it is larger than $f_{\lceil a\rceil}(p)$.
\end{corollary}
\begin{proof}
Take $t=a$ and $\alpha=c_a(s)$ in Theorem~\ref{thm:general}. For $d\le\lfloor a\rfloor$ we have
$\alpha d\le g(d)$, because $g(d)/d$ is non-increasing. For $d\ge\lceil a\rceil$ we have $\alpha
a\le g(\lceil a\rceil)\le g(d)$. For non-integer $a$, both terms of the minimum exceed
$g(\lceil a\rceil)/\lceil a\rceil$, the first by strict concavity of~$g$.
\end{proof}

For integer densities, no improvement over Theorem~\ref{thm:level} is possible without excluding
more structure.

\begin{proposition}\label{prop:kkn}
Let $k\ge1$. For every $p$, on $K_{k,n}$ with unit weights, the expected fraction of the edges that
become internal is at most $f_k(p)+1/n$. Hence, on every class that contains $K_{k,n}$ for all $n$,
no analysis of the random-centre rule gives a per-level fraction above $\max_pf_k(p)$. Such classes
include the graphs of treewidth at most~$k$ and, for $k\le5$, the $K_{k+2}$-minor-free graphs, both
of density~$k$~\cite{M68}.
\end{proposition}
\begin{proof}
The leaves of a star in a bipartite graph lie on the side opposite to its centre and are pairwise
non-adjacent, so the internal edges are exactly the join edges. A vertex of degree $k$ has a join
edge exactly when it is not a centre and one of the $k$ hubs is, which has probability $s(1-s^k)$.
The hubs have at most $k$ join edges, and $|E|=kn$. Finally, $K_{k,n}$ has treewidth~$k$, and
$K_{k+2}$ has treewidth $k+1$.
\end{proof}

By Proposition~\ref{prop:kkn}, Theorem~\ref{thm:level} is optimal for every integer row of Table~\ref{tab:general} whose class
contains all $K_{a,n}$. In particular, $K_5$-minor-free graphs have the same density as planar
graphs, but the rule cannot beat the exponent $6.41$ on them. The improvement to $5.83$ for planar
graphs comes from excluding $K_{3,n}$, which is what the bipartite bound $b=2$ records.

\paragraph{Surfaces.} A simple graph with $n\ge3$ vertices that embeds in a surface of Euler
genus~$e$ has at most $3n-6+3e$ edges, and at most $2n-4+2e$ if it is bipartite. For $e\le2$ (the
sphere, projective plane, torus and Klein bottle) this gives $a=3$ and $b=2$. Hence
Theorem~\ref{thm:planar} and all its numbers, including the exponent $5.89$ for $p=3/8$, hold for
graphs on these surfaces. For larger $e$ we use
$a_e=\max\{3,\max_n\min\{(n-1)/2,\,3-(6-3e)/n\}\}$ and
$b_e=\max\{2,\max_n\min\{\lfloor n^2/4\rfloor/n,\,2-(4-2e)/n\}\}$. Table~\ref{tab:surfaces} lists
the resulting exponents. Every entry of the last column uses a certificate for
Theorem~\ref{thm:general} that was found by linear programming (\texttt{sim/tight/surfaces.py}). The
certificates and the exponent bounds are checked in Lean (Section~\ref{sec:lean}).

\begin{table}[ht]\centering
\begin{tabular}{rrr|rrrr}\toprule
Euler genus $e$ & $a_e$ & $b_e$ & \cite{AS26} & Thm.~\ref{thm:level} & Cor.~\ref{cor:aonly} &
Thm.~\ref{thm:general}\\\midrule
$0,1,2$ & $3$ & $2$ & 25.81 & 6.42 & 6.42 & 5.83\\
3 & $27/8$ & $20/9$ & 29.11 & 7.66 & 6.42 & 6.13\\
4 & $11/3$ & $12/5$ & 31.67 & 7.66 & 6.97 & 6.55\\
6 & $21/5$ & $30/11$ & 36.36 & 8.88 & 7.66 & 7.19\\
8 & $51/11$ & $3$ & 40.20 & 8.88 & 8.19 & 7.75\\
12 & $11/2$ & $24/7$ & 47.79 & 10.08 & 9.19 & 8.76\\
20 & $33/5$ & $70/17$ & 57.46 & 11.28 & 10.60 & 10.10\\
100 & $27/2$ & $8$ & 118.10 & 19.45 & 18.73 & 18.17\\
200 & $37/2$ & $11$ & 162.04 & 25.18 & 24.50 & 23.93\\\bottomrule
\end{tabular}
\caption{Round exponents $\ln3/\ln(1/q)$ for graphs on a surface of Euler genus $e$ (for an
orientable surface, $e$ is twice the genus), at the best $p$ with at most $8$ fair coins. Our
exponents are rounded up.}
\label{tab:surfaces}
\end{table}

\section{Plugging the rule into the clustering algorithm}\label{sec:integration}

We refer to the numbering of~\cite{AS26} and recall its notation. The clustering acts on a weighted
graph $H$ of the class $\mathcal C$ and maintains partitions $\mathcal P_0,\mathcal P_1,\dots$ of $V(H)$
into connected parts, each with a leader. For a partition $\mathcal P$, $\cut_w(\mathcal P)$ is the
total weight of the edges of $H$ between different parts, and $H/\mathcal P$ is the quotient whose
vertices are the parts and whose edge weights add up the weights between two parts. The
$\sigma$-algebra $\mathcal F_{t-1}$ is generated by the coins of the levels before~$t$, and
$T_{\mathcal C}$ is as in~\cite[Thm.~5.4]{AS26}. In Algorithm~1 (\textsc{RandomCluster}), we replace lines 6--13 at
level $t$ as follows. Every part $P\in\mathcal{P}_{t-1}$ reads $b$ level-$t$ coins of its leader and
becomes a centre with probability $p=m/2^b$. Every non-centre part with a centre neighbour in
the quotient $Q=H/\mathcal{P}_{t-1}$ joins its heaviest centre neighbour, with ties broken by
identifier. Every centre absorbs the parts that joined it, and every other part is unchanged.

\begin{itemize}
\item \emph{Lemma 4.1} (partition, radius $r_t=3r_{t-1}+1$, quotient in $\mathcal{C}$, dependence
  on coins). The proof only uses that the new parts are vertex-disjoint stars whose leaves are
  adjacent to their centre in $Q$, which holds by Lemma~\ref{lem:distinct}.
\item \emph{Lemmas 4.2 and 4.3.} Fix the coins of levels $<t$. Then $Q$ is fixed and lies in
  $\mathcal{C}$, so it is $a$-sparse and has an orientation with out-degree at most $\lceil a\rceil$.
  The centre events are independent, because distinct parts have distinct leaders. By
  Lemma~\ref{lem:distinct} every join edge becomes internal, so
  $\cut_w(\mathcal{P}_t)\le w(Q)-\mathrm{gain}(C)$. Theorem~\ref{thm:level} gives
  $\E[\cut_w(\mathcal{P}_t)\mid\mathcal{F}_{t-1}]\le(1-f_{\lceil a\rceil}(p))\,\cut_w(\mathcal{P}_{t-1})$.
  This replaces Lemma~4.3 with $q'=1-f_{\lceil a\rceil}(p)$, and Corollary~4.4 follows verbatim.
  For planar graphs, $Q$ is a minor of a planar graph and has no parallel edges, so
  Theorem~\ref{thm:planar} gives $q'=1-\varphi(p)$ instead, and for the classes of
  Section~\ref{sec:classes} Theorem~\ref{thm:general} gives $q'=1-sc$. Lemma~4.2 is no longer needed.
\item \emph{Lemma 4.5} (view radius $\rho_t=2(3^t-1)$). Let $P=\mathcal{P}_{t-1}(y)$. The new part
  of $y$ is determined by the centre status of $P$, its quotient neighbours $C$ and their quotient
  neighbours $R\subseteq B_H(y,6r+2)$, together with the join choice $c(R)$ of each such $R$. The
  choice $c(R)$ depends on the quotient edges at $R$ and on the coins of the parts meeting
  $B_H(y,6r+3)$. The proof in~\cite{AS26} shows that the $\rho_t$-view of $y$ reveals all of
  these parts and their level-$t$ coins. Hence $\rho_t$ is unchanged.
\item \emph{Theorem 5.4.} The bound $T_{\mathcal C}+9\cdot3^L$ on the number of rounds is unchanged. The algorithm uses
  $bL$ instead of $L$ fair coins per vertex, and the factor $q$ becomes $q'$ in both expectation
  bounds.
\item \emph{Theorem 6.1} (the derandomization of~\cite{K05}) requires only a prescribed finite
  number of fair coins per vertex, which still holds.
\item \emph{Theorem 6.2 and Corollary 8.1} hold with $q'$ in place of $q$. For planar graphs,
  $p=3/8$ gives $q'=1740037/2097152$, $p=1/2$ gives $q'=977/1152$, and $p=21/64$ gives
  $q'=3642429005643/2^{42}$. This proves
  Theorem~\ref{thm:main}.
\end{itemize}
The same substitution improves the randomized $(1+\varepsilon)$-approximation and the
deterministic fractional $(1+\varepsilon)$-approximation of~\cite[Thm.~6.2]{AS26} for every
proper minor-closed class.

\section{Limits of the approach}\label{sec:limits}

Theorem~\ref{thm:planar} bounds only the join edges, and this cannot be pushed much further.
For unit weights, Lemma~\ref{lem:cake} gives the exact value $\E[\mathrm{gain}]/w=s\sum_vg(d_v)/|E|$.
On the Apollonian network obtained from $K_4$ by inserting a vertex into every face ten times
($118100$ vertices, script \texttt{apo.py}), this value is at most $0.17402$ for every~$p$ (the maximum is at $p\approx0.320$),
and it equals $0.17161$ at $p=3/8$, against $\varphi(3/8)=0.17029$. Deeper networks lower these
values slightly. Hence no lower bound on the expected gain of the rule, for any $p$, exceeds
$0.17402$, and the exponent of any analysis that counts only join edges is at least $5.74$.
Theorem~\ref{thm:planar} gives $5.83$.

Counting the edges between leaves of the same star helps only on graphs with triangles.
Proposition~\ref{prop:kkn} with $k=2$ bounds the rule itself, with every internal edge counted:
planar graphs contain all $K_{2,n}$, so no analysis of the random-centre rule yields a per-level
fraction above $\max_pf_2(p)=1/(3\sqrt3)\approx0.1925$ on planar graphs, that is, an exponent below
$5.13$.

No star contraction of the square grid contracts more than a $2/5+o(1)$ fraction of its
edges. The grid is triangle-free, so a star with $\ell$ leaves internalizes exactly $\ell$ edges.
Each centre has at most $4$ leaves, so at most $4n/5$ edges become internal. Any analysis that
bounds every level by a uniform worst-case fraction, with radius growth $r\mapsto3r+1$, therefore
cannot give an exponent below $\ln3/\ln(5/3)\approx2.15$.

The two remaining propositions bound every analysis that counts only join edges, as ours does.

\begin{proposition}\label{prop:jointop}
For every even $k\ge2$ there is a triangulation $G$ with $n=4k$ vertices and $\gamma(G)=n/4$. In
every partition of $G$ into stars, at most $3n/4$ edges are join edges. Hence, for any one-level
star contraction, an analysis that counts only join edges gives at most a $3n/(4(3n-6))\to1/4$
fraction per level, and an exponent of at least $\ln3/\ln(4/3)\approx3.819$.
\end{proposition}
\begin{proof}
Let $H$ consist of $k$ nested triangles $a_ib_ic_i$, $0\le i<k$, where consecutive triangles are joined by
the six edges $a_ib_{i+1}$, $a_ic_{i+1}$, $b_ia_{i+1}$, $b_ic_{i+1}$, $c_ia_{i+1}$, $c_ib_{i+1}$.
For even $i<k$, the faces $a_ib_ic_{i+1}$ and $a_{i+1}b_{i+1}c_i$ lie in the band between the
triangles $i$ and $i+1$. These $k$ faces are pairwise disjoint and cover $V(H)$. Insert a new vertex
into each of them. The result $G$ is a triangulation with $3k+k=n$ vertices. The closed
neighbourhoods of the inserted vertices partition $V(G)$, so $\gamma(G)\ge k$, and the inserted
vertices form a dominating set. In a partition into stars, the centres and the singleton parts
dominate $G$, so there are at least $\gamma(G)$ parts. Every vertex that is not a centre or a
singleton has exactly one join edge, so there are at most $n-\gamma(G)=3n/4$ join edges, while
$|E(G)|=3n-6$.
\end{proof}

It was conjectured in~\cite{MT96} that every sufficiently large triangulation satisfies $\gamma\le n/4$.
If the conjecture holds, then Proposition~\ref{prop:jointop} is the best such bound for triangulations.

\begin{proposition}\label{prop:indep}
Consider any rule in which every part becomes a centre independently with a probability that
depends only on its weighted radius-$r$ neighbourhood, without identifiers, and every non-centre
joins a centre neighbour. On the $m\times m$ triangular grid with unit weights, the expected number
of join edges is at most $m^2\max_{s}s(1-s^6)+O(rm)$. Hence an analysis of such a rule that counts
only join edges gives at most $2\cdot7^{-7/6}+o(1)\approx0.2066$ per level, and an exponent of at
least $4.74$.
\end{proposition}
\begin{proof}
All vertices at distance more than $r+1$ from the boundary have the same probability~$p$, and so do
their six neighbours, because their neighbourhoods are isomorphic. Such a vertex has a join edge
with probability $s(1-s^6)$, where $s=1-p$. Every vertex has at most one join edge, at most $O(rm)$
vertices lie near the boundary, and the grid has $3m^2-O(m)$ edges. The maximum of $s(1-s^6)/3$ is
attained at $s=7^{-1/6}$.
\end{proof}

\section{The dependence on \texorpdfstring{$\varepsilon$}{epsilon}}\label{sec:lower}

The deterministic $(1+\varepsilon)$-approximation of $\gamma_f$ and the randomized algorithm with
expected size $(1+\varepsilon)\gamma$ run in $(c_0/\varepsilon)^{\kappa}$ rounds, both
in~\cite{AS26} and with the rule of Section~\ref{sec:rule}. Open Problem~3 of~\cite{AS26} asks for the optimal
dependence on~$\varepsilon$. For integral solutions on general graphs, a
$(1+\varepsilon)$-approximation of $\gamma$ needs $\Omega(\log n/\varepsilon)$ rounds~\cite{CL23}.
On every proper minor-closed class, the randomized algorithm of~\cite{AS26} with expected size
$(1+\varepsilon)\gamma$ needs a number of rounds that depends only on~$\varepsilon$.

\subsection{A lower bound on paths}\label{sec:paths}

Paths give a lower bound that is linear in $1/\varepsilon$. We let a vertex learn in $r$ rounds
its radius-$r$ ball together with all identifiers and degrees in it. This only strengthens the
algorithm.

\begin{theorem}\label{thm:lower}
Let $r\ge0$ and $L=2r+6+(r\bmod 3)$. Let $\mathcal A$ be an $r$-round LOCAL algorithm that is
either deterministic and outputs a fractional dominating set, or randomized and always outputs a
dominating set. Then for some identifier assignment on the path $P_L$, the (expected) size of the
output is at least $\gamma(P_L)+2/3=(1+2/L)\gamma_f(P_L)$. This holds even if $\mathcal A$ knows the
graph. Consequently, every $r$-round algorithm of either kind has ratio at least $1+1/(r+4)$ on paths,
and a $(1+\varepsilon)$-approximation needs at least $1/\varepsilon-4$ rounds.
\end{theorem}
\begin{proof}
Let $R=r+1$ and $M=L-2R=4+(r\bmod3)$, and number the vertices $0,\dots,L-1$. Draw the identifiers
as a uniformly random injection into the finite identifier space, independently of the coins of
$\mathcal A$, and let $x_v$ be the expected value of the output at~$v$. Every outcome is a
fractional dominating set, so $x$ is one, and $\sum_vx_v$ is the average of the (expected) output
size over the identifier assignments. If an isomorphism from the radius-$r$ ball around~$u$ to the one around~$v$
maps $u$ to~$v$ and preserves the degrees in~$P_L$, then $x_u=x_v$. Indeed, the output at a vertex
depends only on its ball with the identifiers, the degrees and the coins in it, and the identifiers
and the coins on the two balls have the same distribution. Hence
$x_v=x_{L-1-v}$ for all~$v$, and all $M$ middle vertices $R\le v<R+M$ have the same value~$\mu$:
their balls are paths with $2r+1$ vertices of degree~$2$. Since $M\ge3$, the closed neighbourhood
of $R+1$ consists of middle vertices, so $\mu\ge1/3$.

Choose $k\in\{0,1,2\}$ and $t\ge0$ with $R+k=3t+2$. The closed neighbourhoods of the vertices
$0,3,\dots,3t$ are disjoint. Their union $\{0,\dots,R+k-1\}$ consists of the left part
$\{0,\dots,R-1\}$ and $k$ middle vertices. Hence $\sum_{v<R}x_v+k\mu\ge t+1=(R+k+1)/3$.
By the symmetry, and since $M-2k\ge0$ and $\mu\ge1/3$,
\[
  \sum_vx_v=2\sum_{v<R}x_v+M\mu\ \ge\ \frac{2(R+k+1)}{3}+(M-2k)\mu\ \ge\ \frac{2R+M+2}{3}=\frac{L+2}{3}.
\]
Finally $3\mid L$. The vertices $1,4,\dots,L-2$ dominate $P_L$, and their closed neighbourhoods are
disjoint, so $\gamma(P_L)=\gamma_f(P_L)=L/3$. The ratio bound follows from $L\le2r+8$.
\end{proof}

The bound holds whether or not the algorithm uses identifiers or knows~$n$. The only obstruction
is that the middle vertices cannot see the ends. For fractional domination the bound is close to
exact. Every fractional dominating set of $P_\ell$ has weight at least $\lceil\ell/3\rceil$. On
$P_\ell$ with $\ell\ge2$, put $x_v=2/3$ at the two ends and $x_v=1/3$ at the other vertices. This is a
fractional dominating set of weight $(\ell+2)/3\le(1+2/\ell)\lceil\ell/3\rceil$, and every vertex
knows its value from its own degree. If $\ell\le2r+2$, every vertex sees the nearer end within
distance~$r$. If it also knows~$\ell$, it can output the average of a fixed minimum dominating set
and its mirror image, which has weight $\lceil\ell/3\rceil$. With the weights $2/3$ and $1/3$ for
$\ell\ge2r+3$, an algorithm that knows the length has ratio at most $1+2/(2r+3)$ for every $r\ge0$.
So for fractional algorithms that know the length, the smallest number of rounds for ratio
$1+\varepsilon$, with $0<\varepsilon\le1$, lies between $1/\varepsilon-4$ and $1/\varepsilon-1/2$.
Without the length, let a vertex output the same average if it sees both ends of a path with at
most $r+1$ vertices, and the weights $2/3$ and $1/3$ otherwise. On a path with at most $r+1$
vertices every vertex sees both ends, and on a longer path no vertex uses the average. So the output
is a fractional dominating set, its ratio is at most $1+2/(r+2)$, and $\lceil2/\varepsilon\rceil-2$
rounds suffice for ratio $1+\varepsilon$. For $r\le15$ we computed the optimal ratios numerically.
They are $1+2/L$ for fractional algorithms that know the length. For those that do not, with
identifiers drawn independently of the length, they are $1+1/(r+2)$ if $r\not\equiv2$ and
$1+1/(r+3/2)$ if $r\equiv2\pmod 3$ (script \texttt{path\_sym.py}).

For randomized algorithms that always output a dominating set, the bound is far from exact when $r$
is small. Let $r=0$, and suppose that three identifiers can each be absent from the output, with
positive probability, at a vertex of degree~$2$. On three consecutive inner vertices of a path with
these identifiers, all three are absent with positive probability, and the middle one is not
dominated. So at most two identifiers can be absent. If the identifier range has at least $\ell+2$
elements, the inner vertices of $P_\ell$ can avoid them, so the output has size at least $\ell-2$,
and the ratio tends to $3$ as $\ell$ grows. For these algorithms, the best upper bound on paths we
know is the $O(1/\varepsilon)$ of Theorem~\ref{thm:forests}.

Theorem~\ref{thm:lower} does not cover deterministic algorithms that output dominating sets. In a
constant number of rounds, they cannot beat ratio~$3$ on paths~\cite[Thm.~8.7]{AS26}.

\begin{corollary}\label{cor:dichotomy}
Let $\mathcal C$ be a minor-closed class. If $\mathcal C$ excludes some path, a minimum dominating
set can be computed deterministically in a constant number of rounds. Otherwise,
a deterministic $(1+\varepsilon)$-approximation of $\gamma_f$ on $\mathcal C$, or a randomized
algorithm that always outputs a dominating set of expected size at most $(1+\varepsilon)\gamma$,
needs at least $1/\varepsilon-4$ rounds.
For planar graphs, the optimal number of rounds lies between $1/\varepsilon-4$ and
$O((c_0/\varepsilon)^{5.83})$.
\end{corollary}
\begin{proof}
If $P_k\notin\mathcal C$, every path in a graph of $\mathcal C$ has fewer than $k$ vertices, so every
component has diameter less than~$k$. In $k$ rounds every vertex sees its component, including all
identifiers, and all vertices of a component agree on its lexicographically first minimum dominating
set. Otherwise $\mathcal C$ contains all paths, since it is closed under subgraphs, and
Theorem~\ref{thm:lower} applies. The upper bound for planar graphs is the six-coin exponent
of Theorem~\ref{thm:main}, applied to the two guarantees of~\cite[Thm.~6.2]{AS26} named above
(Section~\ref{sec:integration}).
\end{proof}

\subsection{Forests: the lower bound is tight}\label{sec:forests}

For forests, which have density $a=1$, Table~\ref{tab:general} gives $(c_0/\varepsilon)^{3.82}$
rounds. We show that $O(1/\varepsilon)$ rounds suffice, even with unbounded degrees.

\begin{theorem}\label{thm:forests}
There is a constant $C$ such that for every $\varepsilon\in(0,1]$ the following hold. A deterministic
LOCAL algorithm computes, on every forest $G$, a fractional dominating set of weight at most
$(1+\varepsilon)\gamma_f(G)$ in $C/\varepsilon$ rounds. A randomized LOCAL algorithm always outputs a
dominating set of expected size at most $(1+\varepsilon)\gamma(G)$ in $C/\varepsilon$ rounds.
\end{theorem}

With Theorem~\ref{thm:lower}, the optimal number of rounds on forests is $\Theta(1/\varepsilon)$,
both for the deterministic fractional and for the randomized $(1+\varepsilon)$-approximation.

\begin{proof}
\emph{Repair.} For $U\subseteq V$ let $\mathrm{opt}(U)$ be the smallest weight of an $x\ge0$
supported on $U$ with $x(N[u])\ge1$ for all $u\in U$. Let $B\subseteq U$ meet $N[u]$ for every
$u\in U$ with a neighbour outside~$U$. For every fractional dominating set $y$ of~$G$,
\begin{equation}\label{eq:repair}
\mathrm{opt}(U)\le y(U)+|B|.
\end{equation}
Indeed, $y$ on $U$ plus weight $1$ on $B$ is feasible. The integral version of \eqref{eq:repair}, for
dominating sets, holds in the same way.

\emph{Core and hairs.} Let $F$ be a forest and $R\ge1$. For an edge $vw$ of $F$, let $h(v,w)$ be the
largest distance from $v$ to a vertex of the component of $F-v$ that contains~$w$. The \emph{core}
$K$ consists of the vertices $v$ with $h(v,w)\ge R$ for at least two neighbours~$w$.
\begin{itemize}
\item[(a)] \emph{$K$ meets every component of $F$ in a connected set.} Let $v,v'\in K$ and let $z$ lie
  on the path between them. One of the two high branches of $v$ avoids~$z$, so the branch of $z$
  towards $v$ has height at least~$R$, and likewise towards~$v'$.
\item[(b)] \emph{Every vertex outside $K$ in a component that meets $K$ has distance at most $R$ to
  $K$.} Let $u_0,\dots,u_k$ be a shortest path from $u_0\notin K$ to $u_k\in K$. The branch of $u_i$
  towards $u_0$ has height at least~$i$. The branch towards $u_k$ has height at least~$R$, because it
  contains a high branch of $u_k$ that avoids $u_{k-1}$. So $u_i\in K$ for $R\le i\le k$, and minimality
  gives $k\le R$. Let $a(u)$ be the nearest core vertex, and let
  $\mathrm{hair}(c)=\{c\}\cup a^{-1}(c)$; by~(a), $a(u)$ is unique. Every neighbour $w$ of a vertex
  $u\in\mathrm{hair}(c)\setminus\{c\}$ lies in $\mathrm{hair}(c)$: if $w\in K$ then $a(u)=w$, and
  otherwise the path from one of $u,w$ to $K$ passes through the other.
\item[(c)] \emph{Components of $F$ that miss $K$ have diameter less than $2R$.} On a path of length at
  least $2R$, the vertex at distance $R$ from one end would lie in~$K$.
\item[(d)] \emph{A core vertex $v$ with at most one core neighbour has a path $v,x_1,\dots,x_R$ with
  $x_1,\dots,x_R\in\mathrm{hair}(v)$.} One of the high branches of $v$ starts at a vertex outside~$K$.
  By~(a), this branch contains no core vertex, so it lies in $\mathrm{hair}(v)$, and it contains a
  vertex at distance $R$ from~$v$.
\end{itemize}
Call a core vertex a \emph{branch vertex} if it has no core neighbour or at least three. The other
core vertices induce paths, the \emph{core paths}. Let $L$ be the number of core vertices with at most
one core neighbour, and let $d_K$ be the degree in $F[K]$. Every core vertex $v$ satisfies
\[
  [v\text{ branch}]+2\le d_K(v)+3\,[d_K(v)\le1],\qquad
  [v\text{ branch}]\,d_K(v)+6\le3d_K(v)+6\,[d_K(v)\le1],
\]
as one checks for $d_K(v)=0$, $1$, $2$ and $d_K(v)\ge3$. Since $F[K]$ is a forest,
$\sum_{v\in K}d_K(v)\le2|K|$. Summing over $K$, there are at most $3L$ branch vertices, and
$\sum_bd_K(b)\le6L$ over the branch vertices~$b$.

\emph{Deterministic version.} Apply this to $F=G$, and let $P$ be the set of vertices of the core
paths. For $u\in P$ and $0\le k\le R$, let $P(u,k)$ be the set of vertices of the core path of $u$ at
distance at most $k$ from~$u$, and let $s(u)\in\{0,1,2\}$ be $2$ minus the number of neighbours of $u$
on its core path. For $X\subseteq K$ let $\mathrm{hair}(X)=\bigcup_{c\in X}\mathrm{hair}(c)$. The
windows are:
\begin{itemize}
\item $\mathrm{hair}(P(u,R))$ for every $u\in P$, with weight $1/(2R+1)$;
\item $\mathrm{hair}(P(e,R-t))$ for every $e\in P$ and $1\le t\le R$, with weight $s(e)/(2R+1)$;
\item $\mathrm{hair}(b)$ for every branch vertex~$b$, with weight~$1$;
\item every component of $G$ that misses~$K$, with weight~$1$.
\end{itemize}
On a core path $p_1\dots p_m$, the windows of the first two kinds are the hairs of the sets
$\{p_j:|j-i|\le R\}$ for $1-R\le i\le m+R$, each with weight $1/(2R+1)$. Those with $i\le0$ are the
windows at $e=p_1$ with $t=1-i$, and those with $i>m$ are the windows at $e=p_m$ with $t=i-m$; if
$m=1$, then $s(p_1)=2$ accounts for both. So every vertex of $P$ lies in windows of total weight
exactly~$1$, and by~(b) so does every vertex of its hair. A vertex of the hair of a branch vertex lies
only in that window, and a vertex of a component that misses $K$ only in that component. So every
vertex lies in windows of total weight exactly~$1$.

Let $\partial U$ be the set of vertices of $U$ with a neighbour outside~$U$. By~(b), the neighbours of
a vertex of $\mathrm{hair}(c)\setminus\{c\}$ lie in $\mathrm{hair}(c)$, and so does every neighbour
of $c$ outside~$K$. Hence a vertex of $\partial\,\mathrm{hair}(X)$ lies in $X$ and has a
core neighbour outside~$X$. For $X=P(u,k)$, it is one of the at most two vertices at distance exactly
$k$ from $u$, or it lies in the set $A$ of vertices of $P$ with a branch neighbour. For $X=\{b\}$ it
is~$b$, and the components have no boundary. Each vertex of $A$ lies in windows of the first two kinds
of total weight~$1$, and all these windows together have weight $(|P|+RS)/(2R+1)$, where
$S=\sum_{e\in P}s(e)$. So
\[
  \sum_U\mathrm{wt}(U)\,|\partial U|\le|A|+\frac{2(|P|+RS)}{2R+1}+3L\le|A|+\frac{|P|}R+S+3L.
\]
Every vertex of $A$ is an end of a core edge at a branch vertex, so $|A|\le6L$. The number of
neighbours of $e\in P$ on its core path is $d_K(e)$ minus its number of branch neighbours. So $s(e)$
is $[d_K(e)=1]$ plus the number of branch neighbours of~$e$, and $S\le L+6L=7L$.

Now let $y$ be a fractional dominating set. For every core vertex $c$ with at most one core
neighbour, take the path $c,x_1,\dots,x_R$ of~(d). By~(b), $N[x_j]\subseteq\mathrm{hair}(c)$. A vertex
$w$ lies in $N[x_j]$ only for the $c$ with $w\in\mathrm{hair}(c)$, and for at most three values of $j$,
since the $x_j$ with $w\in N[x_j]$ are at distance at most $2$ from each other. So
$RL\le\sum_{c,j}y(N[x_j])\le3y(V)$. Let $I$ be the set of vertices of $P$ with two core neighbours and
no branch neighbour. A vertex $w\notin K$ lies in $N[z]$ with $z\in I$ only for $z=a(w)$, a branch
vertex lies in no such $N[z]$, and a vertex of $P$ lies in $N[z]$ only for itself and its at most two
neighbours on its core path. So $|I|\le\sum_{z\in I}y(N[z])\le3y(V)$, and
$|P|\le|I|+L+|A|\le3y(V)+7L$. Altogether, since $L\le3y(V)/R$ and $R\ge1$,
\[
  \sum_U\mathrm{wt}(U)\,|\partial U|\le16L+\frac{3y(V)+7L}{R}\le\frac{48+3+21}{R}\,y(V)
  =\frac{72}{R}\,y(V).
\]

Let $x^U$ be the optimal solution of the problem defining $\mathrm{opt}(U)$ that minimizes
$\sum_vx_v^2$; it is unique, so the vertices of $U$ agree on it. Output $x=\sum_U\mathrm{wt}(U)x^U$. A
vertex $u$ is dominated in each window that contains it, and these windows have total weight~$1$. By
\eqref{eq:repair} with $B=\partial U$, for every fractional dominating set~$y$,
\[
  \sum_vx_v\le\sum_U\mathrm{wt}(U)\big(y(U)+|\partial U|\big)\le\Big(1+\frac{72}{R}\Big)y(V).
\]
Take $y$ optimal and $R=\lceil72/\varepsilon\rceil$. A vertex decides membership in $K$ within radius
$R$, membership in $P$ within $R+1$, the value $s(\cdot)$ within $R+2$, and its nearest core vertex
within~$2R$. A window that contains $v$ has its centre within $2R$ of $v$ and all its vertices within
$4R$ of~$v$, by~(b) and~(c). So $v$ finds the windows that contain it, and their problems, in its ball
of radius $7R$, and the algorithm runs in $7R\le511/\varepsilon$ rounds.

\emph{Randomized version.} In $O(1)$ rounds, compute a dominating set $D_0$ with
$|D_0|\le c\gamma(G)$ for an absolute constant~$c$~\cite{LOW08}. Every vertex $u\notin D_0$ picks a
neighbour $d(u)\in D_0$, and $d(v)=v$ for $v\in D_0$. Contracting the stars $S_d=d^{-1}(d)$ gives a
forest $H$ on $D_0$, and a round on $H$ takes $O(1)$ rounds on~$G$. Let $N=|D_0|$. For
$W\subseteq V(H)$ let $V_W=\bigcup_{d\in W}S_d$, and let $\partial W$ be the set of $d\in W$ with an
$H$-neighbour outside~$W$. If $u\in V_W$ has a neighbour $w\notin V_W$, then $d(u)d(w)$ is an $H$-edge
leaving~$W$, so $d(u)\in\partial W\cap N[u]$. So \eqref{eq:repair} applies to $U=V_W$ and
$B=\partial W$.

Take the core and the hairs of $F=H$. By~(d), the hair of a core vertex with at most one
core neighbour has more than $R$ vertices. The hairs are disjoint, so $L\le N/R$. Cut the core edges at
branch vertices; there are at most $6N/R$ of them. Every vertex of a core path draws an independent
uniform value in $[0,1]$, given by the binary expansion of a sequence of its coins. Ties have
probability~$0$, and we break them by the identifiers. For a vertex $u$ of a core path, let $B(u)$ be the set of vertices of its
core path at distance at most $R$ from~$u$, and let the \emph{anchor} of $u$ be the vertex of $B(u)$
with the smallest value. Cut every edge $uv$ of a core path whose ends have different anchors. This is
a one-dimensional low-diameter decomposition in the spirit of~\cite{MPX13}. The clusters are the
resulting components of $H[K]$ with their hairs, together with the components that miss~$K$. All core
vertices of a cluster have the same anchor, which has distance at most $R$ from each of them. By~(b),
every vertex of a hair has distance at most $R$ from its core vertex, and by~(c), the components that
miss $K$ have diameter less than $2R$. So every cluster lies within distance $2R$ in~$H$ of one
vertex: the common anchor of its core vertices, its branch vertex, or any of its vertices. If
$B(u)\ne B(v)$, then $B(u)\cup B(v)$ has at least $R+2$ vertices, and $B(u)\triangle B(v)$ has at most
two. The anchors differ only if the smallest value in $B(u)\cup B(v)$ lies in $B(u)\triangle B(v)$, so
$uv$ is cut with probability at most $2/(R+2)$. The core paths have fewer than $N$ edges, so the
expected number of cut edges is at most $8N/R$.
Each cluster $W$ outputs the lexicographically first smallest set $D_W\subseteq V_W$ that
dominates every vertex of~$V_W$. Every cut edge puts at most two vertices into the sets $\partial W$.
By the integral form of~\eqref{eq:repair}, the expected size is at most
$\gamma+16N/R\le(1+\varepsilon)\gamma$ for $R=\lceil16c/\varepsilon\rceil$. In~$H$, a vertex decides
membership in $K$ and in $P$ within radius $R+1$, and its nearest core vertex within $2R$, as in the
deterministic version. So the anchor of a vertex $u$ of a core path is determined by the ball of
radius $2R+1$ around~$u$, and whether an edge $uw$ of a core path is cut, by the ball of radius
$2R+2$ around~$u$. The cluster of a vertex $d$ of~$H$ lies within distance $4R$ of~$d$, so $d$
determines it, and the edges that leave it, from its ball of radius $6R+2$ in~$H$. A round on $H$
takes $O(1)$ rounds on~$G$, so with the computation of $D_0$, the algorithm runs in
$O(R)=O(1/\varepsilon)$ rounds.
\end{proof}

The numerical checks are in \texttt{tree\_frac.py} and \texttt{tree\_ldd.py}. The proof uses
forests only to build windows of radius $O(R)$ and weighted boundary $O(\gamma_f/R)$ locally. For planar graphs,
such a local family of windows, with no dependence on $n$ or on the maximum degree, would give
$\Theta(1/\varepsilon)$ for the fractional approximation in the same way, and polylogarithmic losses would give
$\tilde O(1/\varepsilon)$. The star contractions of Section~\ref{sec:rule} give radius
$(c_0/\varepsilon)^{\kappa}$.

\subsection{Outerplanar graphs}\label{sec:outerplanar}

On outerplanar graphs, the inner faces take the place of the vertices of a tree. Short faces are
seen locally, long faces are not. We remove the long faces by passing to a covering graph in which
the face structure is a tree, and run the construction of Theorem~\ref{thm:forests} there. Vertices
of high degree are handled by the way core paths are measured.

\begin{theorem}\label{thm:outerplanar}
There is a constant $C$ such that for every $\varepsilon\in(0,1]$ the following hold on every
outerplanar graph~$G$. A deterministic LOCAL algorithm computes a fractional dominating set of
weight at most $(1+\varepsilon)\gamma_f(G)$ in $C/\varepsilon$ rounds. A randomized LOCAL algorithm
always outputs a dominating set of expected size at most $(1+\varepsilon)\gamma(G)$ in
$C/\varepsilon$ rounds.
\end{theorem}

With Theorem~\ref{thm:lower}, the optimal number of rounds for both $(1+\varepsilon)$-approximations
on outerplanar graphs is therefore $\Theta(1/\varepsilon)$, with no dependence on the maximum
degree. On arbitrary graphs of maximum degree~$\Delta$, a deterministic $(1+\varepsilon)$-approximation
of $\gamma_f$ takes $O(\varepsilon^{-4}\log^2\Delta)$ rounds~\cite{KMW06}. In a constant number of
rounds, the best ratio of a deterministic algorithm that outputs a dominating set of an outerplanar
graph is~$5$~\cite{BCGW21}.

\begin{proof}
Let $n=|V(G)|$ and $m=|E(G)|$, and let $y$ be an optimal fractional dominating set of~$G$. As in the
deterministic part of the proof of Theorem~\ref{thm:forests}, we work in $G$ itself, here for both
versions. We repeatedly use the \emph{packing bound}: if the closed
neighbourhoods of the vertices of a set $A$ are pairwise disjoint, then $\gamma_f(G)\ge|A|$, since
each of them has $y$-weight at least~$1$.

\emph{Faces.} Fix an outerplanar embedding. Every vertex lies on the outer face, so the interior of a
cycle $C$ contains no vertex. It is cut by the chords of $C$ into inner faces, all of whose vertices
lie on~$C$, and these faces form a tree under sharing an edge. A chord of the boundary of an inner
face would enclose a vertex. Hence the inner faces are exactly the induced cycles of $G$,
independently of the embedding. They are also isometric: a part of $G$ outside an inner face $F$ is
attached to $F$ at a vertex or at the two ends of an edge, so a path that leaves $F$ returns through
the vertex or edge by which it left. Two inner faces share at most one edge, and two faces with a
common edge share no further vertex, since the common chord separates their other vertices. Fix
$\ell\ge14$. An inner face is \emph{short} if it has at most $\ell$ vertices and a \emph{hole}
otherwise. Short faces are recognized in $\ell$ rounds. An \emph{$\ell$-block} is a class of short faces
under the equivalence relation generated by sharing an edge, together with the vertices and edges
of its faces. An edge on no short face is \emph{thin}. The $\ell$-blocks and the thin edges are the
\emph{pieces}, and every edge lies in exactly one piece. The faces of an $\ell$-block $B$ that
contain a vertex $v$ form a path in the weak dual of~$B$, since the weak dual of the block of $G$ that
contains them is a tree. By the same uniqueness, two short faces at~$v$ lie in one $\ell$-block if
and only if they are joined by a sequence of short faces at~$v$ in which consecutive faces share an
edge at~$v$. So the pieces at $v$ are determined by the faces at~$v$, which have length at most $\ell$
or are seen to be longer within distance~$\ell$. Finally, every cycle $C$ of length at most~$\ell$
lies in one $\ell$-block:
the faces inside $C$ are short, they are connected through shared edges, and every edge of $C$ lies
on one of them.

\emph{The decomposition graph.} Let $D$ be the graph with the following nodes and bags: every short
face, with its vertex set as bag; every thin edge, with its two ends; every vertex $v$ that lies in
no piece or in at least two pieces, with bag~$\{v\}$. Two faces are adjacent if they share an edge. A
vertex node $v$ is adjacent to the thin edges at~$v$ and, for every $\ell$-block $B$ at~$v$, to one
face of $B$ at~$v$, chosen by the identifiers. The bags of two adjacent nodes meet in their
\emph{adhesion}, which is one vertex or one edge. Every edge of $G$ lies in a bag. For every vertex~$v$,
the nodes whose bags contain $v$ form a finite tree~$D_v$. Let $f$ be the
number of inner faces, $f_s$ the number of short faces, $t$ the number of thin edges, and let $p_v$ be
the number of pieces at~$v$. An $\ell$-block $B$ with $n_B$ vertices, $m_B$ edges and $f_B$ faces
satisfies $n_B-m_B+f_B=1$ and contains $f_B-1$ adjacencies of $D$. Hence
$\sum_P(|V(P)|-1)=m-f_s$ over the pieces~$P$. Moreover $\sum_v(p_v-1)=\sum_P|V(P)|-n$, and vertices
with $p_v=1$ are not nodes. Together with $n-m+f=c(G)$, where $c$ counts components, and
$c(D)=c(G)$, this gives
\[
  |E(D)|-|V(D)|+c(D)=\sum_P(|V(P)|-1)-n+c(G)=m-f_s-n+c(G)=f-f_s.
\]
So the cycle rank of $D$ equals the number $h$ of holes.

\emph{Few holes.} We show $h\le3\gamma_f/(\ell-13)$. The inner faces, the bridges and the cut vertices
of $G$ form a forest~$\mathcal T$: two faces are adjacent if they share an edge, and a cut vertex is
adjacent to the bridges at it and to one face of every other block at it. For distinct holes $F$ and
$F'$ in one component, the path from $F$ to $F'$ in $\mathcal T$ leaves $F$ through an edge or a
vertex $s$ of~$F$: a chord of a block, with $F$ and $F'$ on different sides, or a cut vertex. So $s$
separates $F-s$ from $F'-s$, and $F\cap F'\subseteq s$. Let $S(F)$ be the set of these separators of
$F$. Let $\mathcal T_h$ be the union of the paths in $\mathcal T$ between holes. Every separator in
$S(F)$ is determined by the node after $F$ on such a path, which is a neighbour of $F$ in~$\mathcal
T_h$. All leaves of the forest $\mathcal T_h$ are holes, so the degrees of the holes in $\mathcal T_h$
sum to at most $2h$, and in particular $\sum_F|S(F)|\le3h$. Let $W(F)$ be the set of vertices of $F$ at distance at least~$2$ from every
element of~$S(F)$. Since $F$ is an induced isometric cycle, $W(F)$ has at least $|F|-4|S(F)|$ vertices,
and it contains at least $(|F|-4|S(F)|-2)/3$ vertices at pairwise distance at least~$3$. Let
$u\in W(F)$ and $u'\in W(F')$ for distinct holes in one component, and let $s$ be the separator of $F$
towards~$F'$. Then $u'\notin s$, since otherwise $u'\in F\cap F'$ would lie in the
separator of $F'$ towards~$F$. So every path from $u$ to $u'$ passes through $s$, and $d(u,u')\ge3$. By
the packing bound, $\gamma_f\ge\sum_F(|F|-4|S(F)|-2)/3\ge h(\ell-13)/3$.

\emph{Unfolding.} The construction below is local, and all bounds in this proof add up over the
components of~$G$, so we may assume that $G$ is connected. Then $D$ is connected. Let $T$ be the
universal cover of $D$, a tree, and let $\Pi$ be its group of deck transformations. It acts freely on
the nodes of $T$, and it is a free group, so no element other than $1$ has finite order. Every node of
$T$ carries the bag of its image in~$D$. Let $\hat G$ be the graph
whose vertices are the pairs $(s,v)$ with $s\in V(T)$ and $v$ in the bag of~$s$, where $(s,v)$ and
$(s',v)$ are identified whenever $ss'$ is an edge of $T$ and $v$ lies in its adhesion. Its edges join
$(s,u)$ and $(s,w)$ whenever $uw$ is an edge of $G$ in the bag of~$s$. Let $\pi(s,v)=v$.
\begin{itemize}
\item[(i)] \emph{$\pi$ is a covering map, and $\Pi$ acts on $\hat G$ freely and transitively on every
  fibre.} The class of $(s,v)$ consists of the pairs $(s',v)$ with $s'$ in the copy of $D_v$ in $T$
  that contains~$s$, because $D_v$ is a tree. Every edge $vw$ of $G$ lies in one piece, and within an
  $\ell$-block in one face or in two adjacent faces of $D_v$. So the class has exactly one neighbour
  over every neighbour of~$v$. The group $\Pi$ permutes the copies of $D_v$ freely and transitively.
\item[(ii)] \emph{$T$ is a tree decomposition of $\hat G$.} Its adhesions have at most two vertices,
  each bag has diameter at most $\ell/2$ in $\hat G$, and every vertex lies in finitely many bags.
  Every adhesion separates $\hat G$, so a shortest path from an adhesion $X$ to a vertex on one
  side of $X$ stays on that side. Likewise, a path between two vertices of a bag that leaves the bag
  returns through the adhesion by which it left. So the bag of a face is an isometric cycle
  of~$\hat G$.
\item[(iii)] \emph{Two distinct lifts of a vertex have distance more than $\ell$.} We claim that every
  closed walk of length at most $\ell$ lifts, from every starting point, to a closed walk. We use
  induction on the length. A backtrack lifts to a backtrack. A closed walk $w_0\dots w_k$ in which
  some vertex occurs twice among $w_0,\dots,w_{k-1}$ splits into two shorter closed walks. Otherwise
  the walk is a cycle, so it lies in an $\ell$-block~$B$. Its lift stays in the copy of $B$ that
  contains the starting point, and this copy maps isomorphically onto~$B$. Now let $\hat v\ne\hat v'$
  lie over~$v$. A path between them of length at most $\ell$ would project to a closed walk whose
  lift from $\hat v$, by unique path lifting, ends at $\hat v'$.
\item[(iv)] \emph{Locality.} Let $r\ge0$, let $\hat v$ be a lift of~$v$, and let $D_r$ be the
  subgraph of $D$ formed by the nodes whose bags meet the ball of radius $r$ around~$v$ and the
  edges whose adhesions meet it. The nodes whose bags contain a vertex~$u$, and the edges of $D$ whose
  adhesions contain~$u$, are determined by the faces of length at most $\ell$ at~$u$, and so by the
  ball of radius $\ell$ around~$u$. So $v$ computes $D_r$ in $r+\ell$ rounds. The graph $D_r$ is
  connected, since every $D_u$ is a tree and every edge of $G$ lies in a bag. Let $T_r$ be the
  component of the preimage of $D_r$ in $T$ that contains the nodes whose bags contain~$\hat v$. It
  is a tree that covers $D_r$, so it is the universal cover of~$D_r$. If $\hat u$ has distance at most
  $r$ from~$\hat v$, then $T_r$ contains all nodes whose bags contain~$\hat u$, since they form a copy
  of $D_u$ and a shortest path from $\hat v$ to $\hat u$ links it to the copy of~$D_v$. An edge of $T$
  between two nodes whose bags meet the ball has an adhesion that meets the ball, since adhesions
  separate, so it lies in $T_r$ as well. Gluing the bags of $T_r$ as in the definition of $\hat G$ therefore gives the ball of
  radius $r$ around~$\hat v$, with the nodes of $T$ whose bags meet it and the edges of $T$ between
  them. Vertices farther away may be split, which does not matter. So $v$ computes all of this in
  $r+\ell$ rounds.
\end{itemize}

\emph{Core and hairs.} All distances are in $\hat G$. For an edge $ab$ of $T$ with adhesion $X$, the
\emph{side} of $ab$ at $b$ is the component of $T-ab$ that contains~$b$. Let $h(a,b)\in\mathbb
N\cup\{\infty\}$ be the supremum of the distances from $X$ to the vertices in its bags. The pair
$(a,b)$ and this side are \emph{high} if $h(a,b)\ge R$. By~(ii), heights are monotone: if the side of
$cd$ at $d$ is contained in the side of $ab$ at~$b$, then $h(a,b)\ge h(c,d)$. Also by~(ii), a shortest
path from $X$ to a vertex on the side at $b$ stays on that side, so $h(a,b)\ge R$ if and only if some
vertex on that side has distance exactly $R$ from~$X$. The core $K$ consists of the nodes $a$ with
$h(a,b)\ge R$ for at least two neighbours~$b$. So whether a node lies in $K$ is determined by the ball
of radius $R+\ell$ around any vertex of its bag, together with the nodes of $T$ whose bags meet that
ball. All of this is invariant under~$\Pi$, which preserves bags and identifiers.
\begin{itemize}
\item[(a)] \emph{$K$ meets every component of $T$ in a connected set.} The proof is that of
  Theorem~\ref{thm:forests}(a). So every node outside $K$ in a component that meets $K$ has a unique
  nearest core node $a(\cdot)$. Let $\mathrm{hair}(c)=\{c\}\cup a^{-1}(c)$. As in
  Theorem~\ref{thm:forests}(b), an edge of $T$ that leaves $\mathrm{hair}(c)$ joins $c$ to a core
  node.
\item[(b)] \emph{Every vertex in a bag of $\mathrm{hair}(c)$ has distance less than $R+\ell/2$ from
  the bag of~$c$.} Let $q$ be the neighbour of $c$ towards a node of $\mathrm{hair}(c)$ whose bag
  contains the vertex. The side of $qc$ that contains $c$ contains a high side of~$c$, so
  $h(q,c)\ge R$. Since $q\notin K$, we get $h(q,q')<R$ for all other neighbours $q'$ of~$q$. Each
  adhesion of $q$ lies in its bag, which has diameter at most $\ell/2$.
\item[(c)] \emph{Components of $T$ that miss $K$ are finite, and their vertices span radius less than
  $R+\ell$.} Along an infinite ray, the adhesions leave every ball by~(ii), so far out both directions
  are high. In a finite component, follow high directions without backtracking. The walk ends at a
  node with no high direction, or at an edge that is high in both directions and has no other high
  direction at its ends.
\item[(d)] \emph{Hairs are disjoint and inject into $G$.} If $\mathrm{hair}(c)$ and
  $\mathrm{hair}(c')$ contain nodes $s$ and $gs$ with $g\in\Pi$, then $gc=c'$. Fix an ownership,
  that is, a node of $D_v$ for every $v$, chosen by the identifiers in the bags of $D_v$, and lift it. Then the vertices owned by the hairs of
  pairwise inequivalent core nodes project to disjoint sets, and injectively.
\item[(e)] \emph{Hairs are local.} Let $z$ be a vertex in the bag of a node $s\notin K$. The component
  of $T-K$ that contains $s$ lies in $\mathrm{hair}(c)$ for its only core neighbour~$c$, or it is a
  component of $T$ that misses~$K$. By~(b) and~(c), all vertices in its bags have distance less than
  $2R+2\ell$ from~$z$. So the hair that contains $s$ is determined by the ball of radius $3R+3\ell$
  around~$z$, together with the nodes of $T$ whose bags meet it.
\end{itemize}

\emph{Runs.} A core node is a \emph{branch node} if it has a number of core neighbours other
than~$2$. The other core nodes have exactly two core neighbours, so the components of the subgraph
of $T$ that they induce are paths, the \emph{core paths}. For a core path $P$, let $\Pi_P$ be the
subgroup of the $g\in\Pi$ with $gP=P$. If $p$ and $gp$ lie on $P$, then $g\in\Pi_P$, since $gP$ is a
core path that meets~$P$. Every element of $\Pi_P$ acts on $P$ as an automorphism. A one-sided
infinite path has no automorphism other than the identity. On a finite or two-sided infinite path,
every automorphism other than a translation fixes a node or exchanges the ends of an edge, and then
its square fixes a node. Since $\Pi$ acts freely on the nodes and has no elements of finite order,
every $g\ne1$ in $\Pi_P$ acts on $P$ as a translation. So if $P$ is finite, then $\Pi_P=1$, and $P$
maps injectively into~$D$. If $P$ is infinite, then it contains nodes $p$ and $gp$ with $g\ne1$, since
$D$ is finite. So $P$ is infinite in both directions, and $\Pi_P$ is infinite cyclic, generated by a
translation by some number $q\ge1$ of nodes. Two nodes of $P$ have the same image in $D$ if and only
if they lie in one orbit of~$\Pi_P$, and we call $q$ consecutive edges of $P$ a \emph{period}. Number
the nodes of a core path as $p_i$, with $i$ in an interval of~$\mathbb Z$, and let $X_i$ be the adhesion of the edge $e_i=p_ip_{i+1}$. For a vertex~$v$, the
indices $i$ with $v\in X_i$ form an interval, because the nodes whose bags contain $v$ form a subtree.
If this interval is nonempty, it is the \emph{run} of~$v$, and its \emph{length} is its number of
indices. Every index lies in the runs of the at most two vertices of its adhesion. A run is
\emph{maximal} if no other run strictly contains it, and every index lies in a maximal run. Two
vertices may have the same run, and we count runs as intervals of indices. Let
$\rho_i$ be the length of the shortest maximal run that contains~$i$. For a node $p$ of a core path,
with adhesions $X$ and $X'$ towards its two core neighbours, let $\lambda(p)=d(X,X')$ if
$X\cap X'=\emptyset$, and $\lambda(p)=0$ otherwise. Give $e_i$ the length
\[
  \mu(e_i)=\frac1{\rho_i}+\frac{\lambda(p_i)+\lambda(p_{i+1})}2,
\]
which does not depend on a direction and is invariant under~$\Pi$. So $\mu$ takes finitely many
values, all positive. A vertex of high degree may lie in the adhesions of many consecutive edges, but
these edges are then short. The bags that contain a vertex $x$ lie within distance $\ell/2$ of~$x$.
So by the locality of~$K$, the runs of the vertices of $X_i$, and with them $\rho_i$ and $\mu(e_i)$,
are determined by the ball of radius $R+2\ell$ around a vertex of~$X_i$, however long these runs are. Let $I_1,I_2,\dots$ be the
maximal runs, ordered by their left ends. None contains another, so
their right ends are ordered in the same way. No index lies in three of them, so
$I_c\cap I_{c+2}=\emptyset$.
\begin{itemize}
\item[(R1)] \emph{If $i\in I_c$ and $j\in I_{c'}$ with $c'\ge c+3$, then $X_i\cap X_j=\emptyset$.} The run
  of a common vertex would contain the right end of $I_c$ and the left end of~$I_{c+3}$. It would lie
  in a maximal run that contains the right end of~$I_c$, that is, in $I_c$ or in~$I_{c+1}$. Neither
  meets~$I_{c+3}$.
\item[(R2)] \emph{If $k$ maximal runs meet an interval $J$ of indices, then
  $(k-4)/2\le\sum_{i\in J}1/\rho_i\le k$.} For the upper bound, $1/\rho_i\le\sum_{I\ni i}1/|I|$ over the
  maximal runs~$I$. For the lower bound, $\sum_{i\in I}1/\rho_i\ge1$ for every maximal run $I$, since
  $\rho_i\le|I|$ for $i\in I$. All but at most four of the $k$ runs lie in $J$, since a run that sticks
  out contains an end of~$J$, and every index lies in at most two of them.
\item[(R3)] \emph{For $a\le b$, $d(X_a,X_b)\le2\sum_{i=a}^b\mu(e_i)+3$.} Let $I_1,\dots,I_k$ be the
  maximal runs that meet $[a,b]$, in order, and let $v_c$ be a vertex with run~$I_c$. Then
  $v_1\in X_a$ and $v_k\in X_b$, since the first of these runs contains $a$ and the last contains~$b$. If $I_c$ and $I_{c+1}$ share an index, then $d(v_c,v_{c+1})\le1$.
  Otherwise $I_c$ ends at some $i$ and $I_{c+1}$ starts at~$i+1$, where $a\le i<i+1\le b$ since both
  runs meet $[a,b]$. Among the maximal runs, only $I_c$
  contains~$i$, so $X_i\cap X_{i+1}=\emptyset$. Then $d(v_c,v_{c+1})\le\lambda(p_{i+1})+2$, and
  $\lambda(p_{i+1})\ge1$ is counted in full in $\mu(e_i)+\mu(e_{i+1})$. Hence
  $d(X_a,X_b)\le k-1+2\sum\lambda(p_{i+1})$, and $k-1\le2\sum_{i=a}^b1/\rho_i+3$ by~(R2).
\end{itemize}

\emph{Windows.} On every core path, fix $\psi$ with $\psi(p_{i+1})=\psi(p_i)+\mu(e_i)$. The windows are:
\begin{itemize}
\item for every core path and every real $\tau$ for which the following set is nonempty, the union
  of $\mathrm{hair}(p)$ over its nodes with $\tau\le\psi(p)\le\tau+R$, with weight $d\tau/R$;
\item $\mathrm{hair}(b)$ for every branch node~$b$, with weight $1$;
\item every component of $T$ that misses $K$, with weight~$1$.
\end{itemize}
Since $\mu$ takes finitely many positive values, in every bounded range of $\tau$ the first family
changes only finitely often, so sums over windows
are finite sums of integrals. These families do not depend on the numbering of the core paths,
neither on the origin nor on the direction, and they are invariant under~$\Pi$. Every node of $T$
lies in windows of total weight exactly~$1$. We identify a window with its node set~$Q$ and with the
set $W$ of vertices owned by~$Q$. If a path window has the core nodes $p_a,\dots,p_b$, then
$\sum_{i=a}^{b-1}\mu(e_i)\le R$. By~(R3), every $X_i$ with $a\le i<b$ lies within distance $2R+3$
of~$X_a$. Every bag of $p_{a+1},\dots,p_b$ contains one of these adhesions and has diameter at most
$\ell/2$. With~(b), every vertex of the window has distance less than $3R+\ell+4$ from a fixed vertex
of $X_a$, or of the bag of $p_a$ if $a=b$. By~(b) and~(c), the other windows have diameter less than
$2R+2\ell$. So every window has diameter less than $6R+2\ell+8$.

\emph{Output.} For a finite $W\subseteq V(\hat G)$ let $\mathrm{opt}^+(W)$ be the smallest weight of an
$x\ge0$ supported on $N[W]$ with $x(N[u])\ge1$ for all $u\in W$, and let $x^W$ be the optimal
solution that minimizes $\sum_vx_v^2$. Then $x^{gW}=g\,x^W$. Vertex $v$ picks a lift $\hat v$ and
outputs $x_v=\sum_W\mathrm{wt}(W)\,x^W_{\hat v}$ over the windows with $\hat v\in N[W]$. The value
does not depend on the lift. Since $\pi$ maps $N[\hat u]$ bijectively onto $N[u]$,
\[
  x(N[u])=\sum_{\hat w\in N[\hat u]}\ \sum_W\mathrm{wt}(W)\,x^W_{\hat w}\ \ge\
  \sum_{W\ni\hat u}\mathrm{wt}(W)\,x^W(N[\hat u])\ \ge\ 1.
\]
If $gQ=Q$ for a window, then $g$ permutes the finite set~$Q$, so $g^k$ fixes a node for some
$k\ge1$, and hence $g^k=1$ and $g=1$. So distinct elements of $\Pi$ map a window to distinct windows,
and the pairs $(v,W)$ with $\hat v\in N[W]$ correspond to the pairs of a $\Pi$-orbit $[W]$ and a
vertex of $N[W]$. Hence
$\sum_vx_v=\sum_{[W]}\mathrm{wt}(W)\,\mathrm{opt}^+(W)$.

\emph{Repair.} Let $\hat y=y\circ\pi$. For a
window with node set $Q$, let $\partial Q$ be the set of edges of $T$ with exactly one end in~$Q$. We
claim $\mathrm{opt}^+(W)\le\hat y(W)+\sum_{e\in\partial Q}|X_e|$, where $X_e$ is the adhesion of~$e$.
Let $z\in W$ have a neighbour $w\notin W$. Walk in $T$ from the owner of $z$ to a node whose bag
contains $zw$ through nodes containing~$z$, and then to the owner of $w$ through nodes containing~$w$.
The first edge of this walk that leaves $Q$ lies in~$\partial Q$, and its adhesion contains $z$ or~$w$.
Both lie in $N[W]$ and dominate~$z$. Pick one of them for every such~$z$. The picked vertices lie in
the adhesions of the edges in~$\partial Q$, and weight $1$ on them together with $\hat y$ on $W$ is
feasible.
Summing over orbits, and using $\sum_{[W]}\mathrm{wt}(W)\,\hat y(W)=\sum_vy_v$, we get
\[
  \sum_vx_v\le\gamma_f(G)+2\sum_{[e]}c(e),
\]
where $c(e)$ is the total weight of the windows with $e\in\partial Q$, and $[e]$ ranges over the
$\Pi$-orbits of edges of $T$, that is, over $E(D)$.

\emph{Counting.} By~(a), $c(e)=0$ unless $e$ joins two core nodes. If $e$ lies on a core
path, then $c(e)\le2\mu(e)/R$. If $e$ meets a branch node, then $c(e)\le2$. The core projects to
a subgraph $K_D$ of $D$, and the number of core neighbours of a node equals its degree in~$K_D$. For
a finite graph, $\sum_{\deg\ge3}(\deg-2)$ is at most twice the cycle rank plus the number of
vertices of degree~$1$. Here $\mathrm{rk}(K_D)\le\mathrm{rk}(D)=h$. So the branch nodes of $D$ have at
most $6h+4N_1$ core neighbours in total, where $N_1$ is the number of core nodes of $D$ with at most
one core neighbour. Such a core node $c$ has a high side whose first node is not in~$K$. By~(a), this
side lies in $\mathrm{hair}(c)$. It contains a shortest path $u_0,\dots,u_R$ from its adhesion, and
the at least $(R-2)/3$ vertices $u_2,u_5,u_8,\dots$ have pairwise distance at least~$3$. Their closed
neighbourhoods consist of vertices owned by $\mathrm{hair}(c)$. By~(d) and the packing bound,
$N_1\le3\gamma_f/(R-2)$. A finite core path ends at two core edges that meet branch nodes. An
infinite one projects onto a component of $K_D$ that is a cycle, since each of its nodes has its two
core neighbours on it, and infinite core paths in distinct orbits give distinct components. So there
are at most $4h+2N_1$ core paths up to~$\Pi$.

Next we bound $\sum\mu(e)$ over the edges $e$ of $D$ on core paths. Fix a core path~$P$, and let
$k_P$ be its number of maximal runs, up to~$\Pi_P$ if $P$ is infinite. As in the proof of~(R2),
$1/\rho_i\le\sum_{I\ni i}1/|I|$. Summed over the edges of $P$, or over one period if $P$ is infinite,
the right side counts every maximal run once, up to~$\Pi_P$. So the terms $1/\rho_i$ sum to at
most~$k_P$. Number its maximal runs $I_0,I_1,\dots$, starting
anywhere if $P$ is infinite, so that a generator of $\Pi_P$ maps $I_c$ to~$I_{c+k_P}$. Pick an index $t_d\in I_d$
for every $d\equiv0\pmod{12}$. For $c\equiv6\pmod{12}$ with $c+6\le k_P-4$, let $x_c$ be a vertex of an
adhesion with index in~$I_c$. By~(R1), the adhesions with indices in $I_c$, $I_{c\pm3}$ and $I_{c\pm6}$
are pairwise disjoint. Every adhesion separates $\hat G$, so every path from $x_c$ to $X_{t_{c\pm6}}$
has an inner vertex in an adhesion with index in $I_{c\pm3}$. Hence $N[x_c]$ consists of vertices
owned by the hairs of the nodes $p_i$ with $t_{c-6}<i\le t_{c+6}$. For distinct $c$ these nodes are
distinct. If $P$ is infinite, they lie in one period, since $t_{c+6}$ lies before the left end of
$I_{k_P-2}$, and thus before the left end of the translate~$I_{k_P}$ of~$I_0$. There are at least
$(k_P-28)/12$ values of~$c$. By~(d) and the packing bound, $\sum_Pk_P\le12\gamma_f+28(4h+2N_1)$, over
the core paths up to~$\Pi$. Now let $p$ be a node of a core path with $\lambda(p)\ge6$. Its bag is a
face, since thin edges have $\lambda\le1$ and vertex nodes have $\lambda=0$, and by~(ii) it is an
isometric cycle. A shortest path in it between the two core adhesions of $p$ contains at least
$(\lambda(p)-5)/3$ vertices at distance at least~$3$ from both adhesions and at pairwise distance at
least~$3$. Their closed neighbourhoods consist of vertices owned by $\mathrm{hair}(p)$. By the
packing bound, $\sum_p\lambda(p)\le3\gamma_f+5\,|\{p:\lambda(p)>0\}|$. A node $p_{i+1}$ with
$\lambda(p_{i+1})>0$ that is not an end of its core path has $X_i\cap X_{i+1}=\emptyset$, so a
maximal run ends at~$i$. So at most $k_P+2$ such nodes lie on $P$, up to~$\Pi$. Altogether,
\[
  \sum_e\mu(e)\le\sum_Pk_P+\sum_p\lambda(p)\le6\sum_Pk_P+3\gamma_f+10(4h+2N_1)
  \le75\gamma_f+178(4h+2N_1).
\]

So $\sum_{[e]}c(e)\le150\gamma_f/R+(1424/R+12)h+(712/R+8)N_1$. Take $\ell=R+13$. For $R\ge1500$, the
bounds on $h$ and $N_1$ give $\sum_{[e]}c(e)\le215\gamma_f/R$, so $\sum_vx_v\le\gamma_f+430\gamma_f/R$.
Take $R=\lceil1500/\varepsilon\rceil$. The windows $W$ with $\hat v\in N[W]$, together with
$N[W]$, lie within distance $6R+2\ell+10$ of~$\hat v$. By~(e) and the locality of $K$ and~$\mu$,
these windows, their weights and the problems $\mathrm{opt}^+(W)$ are determined by the ball of radius
$9R+5\ell+10$ around~$\hat v$. By~(iv), $v$ computes $x_v$ in $9R+6\ell+10=15R+88\le22603/\varepsilon$
rounds.

\emph{Randomized version.} Let $\ell=6R+5$. Cut every core edge at a branch node, and every edge $e$
of a core path with $\mu(e)>R$. This splits the core paths into \emph{segments}. Every edge $e$ of a
segment gets an exponential clock of rate $\mu(e)$, that is, the value $-\ln U/\mu(e)$ for a uniform
value $U\in[0,1]$ given by the binary expansion of a sequence of coins. There is one clock for each
edge of $D$, drawn by the vertex with the smallest identifier in its adhesion, so the clocks are
invariant under~$\Pi$. Ties are broken by the identifier of the drawing vertex and then by the order
in which it assigns its coins to its edges. For
a node $p$ of a segment, let $B(p)$ be the set of edges of the segment whose ends have $\psi$-distance
at most $R$ from~$p$. If $p$ has an edge in the segment, this edge lies in $B(p)$, and the
\emph{anchor} of $p$ is the edge of $B(p)$ with the smallest clock. Cut every edge of a segment whose
ends have different anchors. The clusters are the components of the remaining core with their hairs,
and the components of $T$ that miss~$K$. All core nodes of a cluster lie within $\psi$-distance $R$
of the ends of a common anchor. Since $\mu$ takes finitely many positive values, a $\psi$-interval of
bounded length contains finitely many nodes, and every cluster is finite. As for the windows, (R3) shows that it has radius
$O(R+\ell)=O(R)$. The clocks, the cuts and the hairs are invariant under~$\Pi$, so $\Pi$ permutes
the clusters, and they partition the nodes of~$T$. If a cluster $C$ and $gC$ share a node, then
$gC=C$, and $g=1$ as for the windows. So $\pi$ is injective on the vertices owned by a cluster, and
the projections of the clusters partition $V(G)$. Each projected cluster $C$ outputs the
lexicographically first smallest set in $N[C]$ that dominates~$C$. The walk argument, applied to a
lift of a boundary vertex of~$C$, bounds its size by $|S\cap C|+\sum_{e\in\partial Q}|X_e|$ for a
minimum dominating set~$S$. So the expected size is at most $\gamma+4\,\E[\#\text{cut edges of }D]$.

Let $e_i$ and $e_{i+k}=ge_i$ with $g\ne1$ and $k>0$ be edges of one core path. By~(iii),
$d(X_i,X_{i+k})>\ell-1$, so $\sum_{j=i}^{i+k}\mu(e_j)>(\ell-4)/2>3R$ by~(R3). Let $e=pp'$ be an edge of a segment, with $\psi(p)<\psi(p')$. The
edges in $B(p)\cup B(p')$ lie in a $\psi$-interval of length $\mu(e)+2R\le3R$, so they are lifts of
distinct edges of $D$, and their clocks are independent. In particular, ties have probability~$0$. The anchors of $p$ and $p'$ differ only if
the smallest clock in $B(p)\cup B(p')$ lies in $B(p)\triangle B(p')$. This happens with probability
$\mu(B(p)\triangle B(p'))/\mu(B(p)\cup B(p'))$, where $\mu$ of a set of edges is their total length.
An edge $f$ whose ends sit at $\psi$-positions $x<y$ lies in $B(q)$ exactly when
$\psi(q)\in[y-R,x+R]$. If $f\in B(p)\setminus B(p')$, then $\psi(p)-R\le x<\psi(p')-R$, and all edges of the
segment between $x$ and $\psi(p')$ lie in~$B(p)$. So $\mu(B(p))>R$ whenever $B(p)\not\subseteq B(p')$,
and symmetrically for~$p'$. Hence $e$ is cut with probability at most $\mu(B(p)\triangle B(p'))/R$.
Moreover,
$f\in B(p)\triangle B(p')$ only if $y-R\in(\psi(p),\psi(p')]$ or $x+R\in[\psi(p),\psi(p'))$. This happens for
at most two edges $e$ of the segment. Summing over the orbits, we bound the expected number of anchor cuts in $D$
by $2\sum_f\mu(f)/R$, over the edges $f$ of $D$ on core paths. At most $\sum_f\mu(f)/R$ of them have
$\mu(f)>R$, and at most $6h+4N_1$ core edges meet a branch node. With the bounds of the counting
and $\ell=6R+5$, at most $242\gamma_f/R$ edges of $D$ are cut in expectation for $R\ge2200$, and the
expected size is at most $\gamma+968\gamma/R$. Take $R=\lceil2200/\varepsilon\rceil$. Whether an edge
of a segment is cut depends on $\mu$ and on the clocks within $\psi$-distance $R$ of its ends. A
cluster spans a $\psi$-interval of length at most $3R$. By~(R3), (b) and~(c), the clusters $C$ with
$\hat v\in N[C]$, and all edges whose clocks decide them, lie within distance $O(R+\ell)$
of~$\hat v$. By~(e), the locality of $K$ and $\mu$, and~(iv), $v$ computes its output in
$O(R+\ell)=O(1/\varepsilon)$ rounds.
\end{proof}

The numerical checks are in \texttt{outer\_frac.py} and \texttt{outer\_cover.py}. The first works
in $G$ itself, on graphs whose holes are longer than all windows: rings, chains of holes and random
outerplanar graphs of maximum degree $3$, $4$ and~$6$, and graphs with vertices of degree up to~$320$,
namely fans, strips of alternating fans, Farey graphs, a hole with large fans glued to its outside,
and random graphs that attach new faces preferentially at vertices of high degree. For
$R\in\{4,8,16\}$, the windows have weight exactly~$1$ at every vertex, the averaged $x$ dominates,
and the excess $\sum_vx_v-\gamma_f$ is at most the repair bound. On these graphs that bound stays
below $19\gamma_f/R$, against $430\gamma_f/R$ in the proof, and it does not grow with the maximum
degree. Every window lies within distance $R+\ell$ of one of its vertices. For the randomized
version, the expected number of long and anchor cuts stays below $1.1\sum_f\mu(f)/R$, against
$3\sum_f\mu(f)/R$ in the proof.

The second script builds $T$ and $\hat G$ and runs both versions on them as in the proof. It checks
the intermediate claims one by one: the faces, the cycle rank of~$D$, the separators and the packing
behind the bound on~$h$, (i)--(iii), (b)--(d), (R1)--(R3), the packing sets behind the bounds on
$N_1$, $\sum_Pk_P$ and $\sum_p\lambda(p)$, the bounds on $c(e)$ and on the diameters of the windows,
and the repair bound for every window. For the randomized version, it computes the cut probabilities
of the anchors exactly and samples the clusters. It runs on $166$ graphs with up to $618$ vertices,
among them windows that wrap around short holes and contain several lifts of a vertex, holes of
length exactly $\ell+1$, vertices of degree up to~$201$, and random choices of the links of vertex
nodes and of the ownership. No check fails. With $\ell=R+13$, holes of length $\ell+1$ and $R\le8$,
the repair bound reaches $23.2\gamma_f/R$. When $\ell$ is much smaller than $R$, as in the tests of
wrapping windows, the bound on $h$ does not apply, and the repair bound reaches $124\gamma_f/R$. With
$\ell=6R+5$ and $R\le3$, the expected number of cut edges of $D$ stays below a third of the bound
$3\sum_f\mu(f)/R+6h+4N_1$ of the proof.

\section{Formal verification}\label{sec:lean}

The Lean files and the scripts named in this paper are in the repository
\begin{center}
  \url{https://github.com/ssenge/MinimumDominatingSetOnPlanarGraphs},
\end{center}
in the directories \texttt{lean/Op4/}, \texttt{sim/tight/} and \texttt{sim/lower/}. This version of
the paper refers to the tag \texttt{arxiv-v1}. The Lean version is \texttt{v4.36.0-rc1}, and Mathlib
is pinned to revision \texttt{d875a8de} in \texttt{lean/lake-manifest.json}.
Every checked theorem depends only on the standard axioms, and no proof is left open.

The file \texttt{RandomCentres.lean} formalizes Section~\ref{sec:rule}, with the random centre set
modelled exactly as independent coins. It checks Lemma~\ref{lem:distinct}
(\texttt{gain\_eq\_join\_edges}),
Lemma~\ref{lem:vertex} (\texttt{vertex\_bound}), Theorem~\ref{thm:level} (\texttt{random\_centres}),
the identity between the arc sum and the edge weight, the value
$f_3(3/8)=1935/12288$ and the bound $f_3\le 63/400$, Lemma~\ref{lem:dominates}, and the bounds
$6.4<\ln3/\ln(1/(1-f_3(3/8)))<6.42$, $\ln3/\ln(1/(1-f_3(1/2)))<7$ and $25<\kappa<26$. The file
\texttt{Orientation.lean} proves the direction of the orientation theorem of~\cite{H65} that we use,
by Hall's theorem: if every vertex set $S$ spans at most $k|S|$ edges, there is an orientation with
out-degree at most~$k$ (\texttt{orientation\_of\_density}). So Theorem~\ref{thm:level} holds for
every such graph (\texttt{random\_centres\_density}).

The file \texttt{Sharper.lean} checks Section~\ref{sec:planar}. Theorem~\ref{thm:planar}
(\texttt{sharper}) is proved in a discrete form of Lemma~\ref{lem:cake}, by induction on the number
of edges of positive weight: the smallest positive weight $t$ is subtracted from every positive
edge, and the gain splits as $t$ times the join indicator plus the gain of the remaining weights.
Also checked are Lemma~\ref{lem:degree} from (A) and (B) (\texttt{degree\_lemma}), the certificates for
$p=3/8$ and $p=1/2$ (\texttt{cert\_three\_eighths}, \texttt{cert\_half}), the values of $\varphi$, and
the bounds $5.88<\kappa'<5.89$ and $\kappa'_{p=1/2}<6.7$. In this file, Lemma~\ref{lem:facts}
enters as a hypothesis on every symmetric weighting whose support lies in that of~$w$. The certificate of
Lemma~\ref{lem:degree} is checked for every $p\in[21/64,1]$ (\texttt{cert\_range}), with $h_1$ and $h_2$
bounded as in the proof. So \texttt{sharper\_range} gives Theorem~\ref{thm:planar} on the whole range,
and \texttt{kappa\_six\_coins} gives $5.82<\kappa'<5.83$ for $p=21/64$.

The file \texttt{General.lean} checks Section~\ref{sec:classes}: Theorem~\ref{thm:general}
(\texttt{general}, for finitely many real thresholds, through the layer-cake core
\texttt{sharper\_core}), its degree inequality (\texttt{degree\_lemma\_gen}), and
Corollary~\ref{cor:aonly} (\texttt{aonly}, with the certificate \texttt{cert\_aonly}). Here too, the
facts of Lemma~\ref{lem:factsgen} enter as hypotheses.

The file \texttt{Density.lean} removes these hypotheses. It proves Lemmas~\ref{lem:facts}
and~\ref{lem:factsgen} from two bounds on the support of~$w$: every vertex set $S$ spans at most
$a|S|$ edges, and for disjoint $S$ and $T$ at most $b(|S|+|T|)$ edges join $S$ to~$T$
(\texttt{facts\_of\_density}, \texttt{planar\_facts}). Both bounds pass to every weighting whose
support lies in that of~$w$. Hence Theorem~\ref{thm:planar} for every $p\in[21/64,1]$
(\texttt{sharper\_range\_density}) and Theorem~\ref{thm:general} (\texttt{general\_density}) need only
these two bounds on~$w$, and Theorem~\ref{thm:level} needs only the first
(\texttt{random\_centres\_densityA}). For planar graphs, $a=3$ and $b=2$ by Euler's formula, which
we take as given.

The file \texttt{Surfaces.lean} checks every row of
Table~\ref{tab:surfaces} with $e\ge3$: the certificate (\texttt{cert\_genus}$e$), the conditions $t\ge a_e$
and $u\ge2a_e$ on its thresholds, the per-level bound under the density bounds $a_e$ and $b_e$,
obtained from \texttt{general\_density} (\texttt{surface\_genus}$e$), and the exponent bound in the
last column (\texttt{kappa\_genus}$e$). The densities $a_e$ and $b_e$ come from Euler's formula on the
surface. The row $e\le2$ is Theorem~\ref{thm:planar}.

The file \texttt{Asymptotic.lean} checks Proposition~\ref{prop:eight}: the limit $8$ for every $p(a)$
with $p\to0$ and $ap\to\infty$ (\texttt{expRatio\_tendsto\_eight}), the limit $8(1-p)$ for a fixed
$p$ (\texttt{expRatio\_tendsto\_const}), and the dyadic choice (\texttt{dyadic\_p}). It also checks that
with $p=1/2$ the exponent is at most that of~\cite{AS26} for every $a\ge1/2$ (\texttt{never\_worse}).

The file \texttt{Table.lean} checks Table~\ref{tab:general}. It is generated from exact rational
data by the script \texttt{lean/scripts/gen\_table.py}. For every row it checks the three
exponents as printed, that the printed $p$ maximizes $f_a$ among all $p=j/256$
(\texttt{best\_dyadic\_}$a$), and the improvement factor. It also checks the six-coin remark after
Theorem~\ref{thm:main}: for every $p\le1$ with $\varphi(p)>0$, the exponent for $p=21/64$ is less
than the exponent for $p$ plus $10^{-3}$ (\texttt{six\_coins\_near\_best}).

The file \texttt{Kkn.lean} checks Proposition~\ref{prop:kkn}. In a graph without triangles, every
edge inside a star is a join edge (\texttt{internal\_subset}). On $K_{k,n}$ with unit weights, for
every $p\in[0,1]$ and every choice among the centre neighbours, the expected number of internal
edges is at most $(f_k(p)+1/n)\,|E|$ (\texttt{kkn\_internal}). So no analysis of the rule gives a
per-level fraction above $f_k(p)$ on a class that contains all $K_{k,n}$ (\texttt{kkn\_optimal}).
For $k=2$ the file also checks $\max_pf_2(p)=1/(3\sqrt3)$ and the exponent bound $5.13$ of
Section~\ref{sec:limits} (\texttt{kkn\_planar}).

The file \texttt{Levels.lean} checks the step from the per-level factor to the number of rounds in
Section~\ref{sec:integration}. If $c_{t+1}\le q\,c_t$ for every $t$ and $x\ge1$, then
$L=\lceil\log x/\log(1/q)\rceil$ levels give $c_L\le c_0/x$ and $3^L\le3x^{\kappa}$ with
$\kappa=\ln3/\ln(1/q)$ (\texttt{levels}). This is how the factor $q'$ enters the exponent of the
round bound $T_{\mathcal C}+9\cdot3^L$ of~\cite[Thm.~5.4]{AS26}. The algorithm of~\cite{AS26} and the
rest of its analysis are not formalized.

The files \texttt{Apollonian.lean}, \texttt{Stars.lean}, \texttt{Nest.lean} and \texttt{TriGrid.lean}
check the other limits of Section~\ref{sec:limits}. That the graphs used there are planar is taken
as given.
\begin{itemize}
\item \texttt{Apollonian.lean} builds the Apollonian network. A face at level $j$ is addressed by a
  face of $K_4$ and a word in $\{0,1,2\}^j$, and its child $k$ replaces corner $k$ by the inserted
  vertex. After ten steps, the vertices of $K_4$ have degree $3\cdot2^{10}$, a vertex inserted at
  level $j$ has degree $3\cdot2^{9-j}$, and there are $118100$ vertices and $354294$ edges. For every
  $p$, the expected join gain with unit weights is at most $0.17402\,|E|$ (\texttt{apo\_join\_le},
  by a polynomial certificate). At $p=3/8$ it is $0.17161\,|E|$ to five decimals, and a per-level
  fraction of at most $0.17402$ gives $\kappa>5.74$.
\item \texttt{Stars.lean} checks the square grid. In a graph without triangles and with maximum
  degree~$\Delta$, a star contraction internalizes at most $\Delta n/(\Delta+1)$ edges. On the
  $m\times m$ grid this is at most $4m^2/5$ of its $2m(m-1)$ edges (\texttt{grid\_internal}), so
  $\kappa>2.15$.
\item \texttt{Nest.lean} checks Proposition~\ref{prop:jointop}. The graph has $n=4k$ vertices,
  $3n-6$ edges and domination number~$k$, and every partition into stars has at most $3k$ join
  edges (\texttt{nest\_join}). Hence the fraction is at most $1/4$, and $\kappa>3.818$.
\item \texttt{TriGrid.lean} checks Proposition~\ref{prop:indep}. A part sees which positions at
  offsets in $[-r,r]^2$ exist. This window contains its radius-$r$ ball, and the part becomes a
  centre independently with a probability that depends only on the window and on~$m$. The expected
  number of join edges on the $m\times m$ triangular grid is at most $0.61974\,m^2+4(r+1)m$
  (\texttt{idfree\_joins}), and the grid has at least $3m^2-12(r+1)m$ edges. So the fraction is at
  most $0.20658$, and $\kappa>4.74$.
\end{itemize}

The file \texttt{PathLower.lean} checks the counting in Theorem~\ref{thm:lower}. Every fractional
dominating set of $P_{2R+M}$ with $M\ge4$ that is constant on the middle and invariant under
reflection has weight at least $(2R+M)/3+2/3$ (\texttt{path\_excess}). For $L=2r+6+(r\bmod3)$, $3$
divides $L$ and $\gamma(P_L)=\gamma_f(P_L)=L/3$ (\texttt{path\_gamma}). So the weight is at least
$(1+2/L)\gamma_f(P_L)$, where $1+2/L\ge1+1/(r+4)$ (\texttt{path\_lower}), and ratio $1+\varepsilon$
forces $r\ge1/\varepsilon-4$ (\texttt{rounds\_lower}).

The files \texttt{LocalPath.lean}, \texttt{LocalRandom.lean} and \texttt{LocalMeasure.lean} check the
symmetrization over identifiers in an explicit model of $r$-round algorithms on~$P_L$. The view of a
vertex lists the identifiers and the degrees of the vertices at offsets $-r,\dots,r$, with an empty
entry for offsets outside the path. In the randomized case it also lists their coins. The coins are
independent of each other and of the identifiers, and all have the same law on an arbitrary
probability space, so real-valued randomness is included. A ball in a path has no
preferred direction, so the algorithm is a function of the view that does not change when the view
is read backwards. It may depend on $L$, $r$ and the size of the identifier space. Suppose its
output is a fractional dominating set for every injective identifier assignment and, in the
randomized case, every outcome of the coins. In the randomized case the output at a vertex must
also be integrable in the coins, which holds if it is measurable and bounded. Then for some
assignment the (expected) weight is at least $(1+2/L)\gamma_f(P_L)$ (\texttt{local\_path\_lower},
\texttt{measure\_path\_lower}). Hence the ratio is at least $1+1/(r+4)$, and ratio $1+\varepsilon$
forces $r\ge1/\varepsilon-4$ (\texttt{local\_rounds\_lower}, \texttt{measure\_rounds\_lower}). As in
the proof of Theorem~\ref{thm:lower}, the identifier assignments and the coins are averaged
together; permuting the coordinates of an independent and identically distributed vector of coins
does not change its law. For coins with finitely many values, \texttt{LocalRandom.lean} gives an
elementary version with finite sums.

The file \texttt{PathUpper.lean} checks the upper bound on paths in the same model. Every
fractional dominating set of $P_L$ has weight at least $\lceil L/3\rceil$
(\texttt{path\_lower\_ceil}). For every $r\ge0$ and $L\ge1$, the algorithm described after the
proof of Theorem~\ref{thm:lower} reads from the view only the distance to the nearer end, capped
at $r+1$. So it does not change when the view is read backwards, and it may use $L$. For every
injective identifier assignment it outputs a fractional dominating set of weight at most
$(1+2/(2r+3))\,y(P_L)$ for every fractional dominating set $y$ of $P_L$
(\texttt{local\_path\_upper}). Hence for $0<\varepsilon\le1$ some $r\le1/\varepsilon-1/2$
gives ratio $1+\varepsilon$ (\texttt{local\_rounds\_upper}).

The file \texttt{Exact.lean} checks the first part of Corollary~\ref{cor:dichotomy}. Every vertex
outputs whether it lies in a fixed minimum dominating set of its component, and the output is a
minimum dominating set (\texttt{exact\_min}). If neither $G$ nor $G'$ has a path on $k$ vertices and
the two graphs agree within distance $k$ of~$v$, the outputs at $v$ are equal
(\texttt{exact\_local}). So $k$ rounds suffice.

The file \texttt{Windows.lean} checks, for finite graphs, the steps of Theorems~\ref{thm:forests}
and~\ref{thm:outerplanar} that do not depend on the graph class: the packing bound
(\texttt{packing}), the repair~\eqref{eq:repair} (\texttt{repair}) and the averaging over windows
(\texttt{window\_average}). For the averaging, let the windows cover every vertex with total weight
exactly~$1$. Let every window $U$ come with a set $B$ that meets $N[u]$ for every $u\in U$ with a
neighbour outside~$U$, and with a solution that dominates $U$ and is optimal among the vectors
supported on a given set that contains $U$ and~$B$. Then the weighted average of these solutions is
a fractional dominating set of weight at most $y(V)+\sum_U\mathrm{wt}(U)\,|B|$ for every fractional
dominating set~$y$. This gives the first inequality of the averaging step in the proof of
Theorem~\ref{thm:forests}. In Theorem~\ref{thm:outerplanar}, \texttt{packing} applies as it is, and
\texttt{repair}, on the finite graph induced by $N[W]$, gives the feasible solution of the repair
step.

The files \texttt{ForestBasic.lean} to \texttt{ForestLocal.lean} check the deterministic part of the
proof of Theorem~\ref{thm:forests} on every finite forest: (a) to~(d), the counts of branch vertices,
the cover by windows of total weight exactly~$1$ (\texttt{cover}, with \texttt{cover\_count} on the
core paths), the bounds $RL\le3y(V)$ (\texttt{card\_L\_le}) and $|P|\le3y(V)+7L$
(\texttt{card\_P\_le}, with the counts of branch vertices), and the bound $(72/R)\,y(V)$ on the weighted boundary
(\texttt{sum\_wt\_wB\_le}). Lean lists the windows at an end $e$ once for each of the $s(e)$
directions, and every component $C$ that misses the core once for each of its vertices, with weight
$1/|C|$. Instead of the optimum of least norm, the output averages one fixed optimal solution of each
window problem, which exists by compactness (\texttt{exists\_isOpt}). It is a fractional dominating
set of weight at most $(1+72/R)\,y(V)$ for every fractional dominating set~$y$
(\texttt{forest\_frac}). On any two forests in which the vertices at distance less than $7R$ from $v$
have the same neighbours, the outputs at $v$ are equal (\texttt{forest\_local}). With
$R=\lceil72/\varepsilon\rceil$ for $0<\varepsilon\le1$, this gives ratio $1+\varepsilon$ in
$7R\le511/\varepsilon$ rounds (\texttt{forest\_frac\_approx}). Here the vertex set
is the identifier space, and the identifiers that are not nodes are isolated vertices; the bound then
holds on the nodes (\texttt{forest\_frac\_on}).

The file \texttt{ForestRand.lean} checks the randomized part on every finite forest and for every
$R\ge1$. The input of~\cite{LOW08} enters as a hypothesis: a set $D_0$ and stars $d$ with
$d(u)\in D_0\cap N[u]$ and $d=\mathrm{id}$ on~$D_0$. The values are independent with any atomless law
on~$\mathbb R$, for instance uniform on $[0,1]$. Ties have probability~$0$ and are broken by a fixed
index, and every cluster outputs a fixed minimum dominating set of~$V_W$. The contraction $H$ is a
forest (\texttt{quot\_acyclic}), adjacent vertices of a core path get different anchors with
probability at most $2/(R+2)$ (\texttt{prob\_anchor\_ne}), and the clusters are those of the proof,
and each lies within distance $2R$ in~$H$ of one vertex (\texttt{cl\_shape},
\texttt{cluster\_radius}). The output always
dominates, and its expected size is at most $|D^*|+16|D_0|/R$ for every dominating set $D^*$
(\texttt{forest\_rand}). Hence $|D_0|\le c\gamma$ and $R\ge16c/\varepsilon$ give expected size at
most $(1+\varepsilon)\gamma$ (\texttt{forest\_rand\_approx}). Whether $v$ is in the output depends
only on the values within distance $15R+1$ of~$v$ (\texttt{randOut\_local\_G}). That the core and
the clusters are computed from balls of radius $O(R)$, and the computation of $D_0$ in $O(1)$
rounds, are proved on paper only.

The file \texttt{Outer.lean} checks the steps of the proof of Theorem~\ref{thm:outerplanar} that can
be stated on their own, with the structure of the covering graph as hypotheses. A core path is
modelled by adhesions $X_i$, each one vertex or one edge, such that for every vertex $v$ the indices
$i$ with $v\in X_i$ form an interval. On such a path, Lean checks (R1), both bounds of (R2) and (R3)
(\texttt{R1}, \texttt{R2\_upper}, \texttt{R2\_lower}, \texttt{R3}). Hence a path window of length at
most $R$ keeps its adhesions within distance $2R+3$ of the first one (\texttt{window\_radius}). It
also checks $\sum_{\deg v\ge3}(\deg v-2)\le2\,\mathrm{rk}+|\{v:\deg v=1\}|$ for finite graphs
(\texttt{branch\_count}), the interval facts of the randomized version on a finite segment
(\texttt{big\_left}, \texttt{big\_right}, \texttt{anchor\_cuts}), the packing on an isometric cycle
(\texttt{cycle\_packing}), and the arithmetic from the per-step bounds to the constants $75$ and
$178$, $215$ and $430$, and $242$ and $968$ (\texttt{sum\_mu\_le}, \texttt{det\_excess},
\texttt{rand\_size}). The faces, the decomposition graph and its cycle rank, the covering graph with
its group action, the core and the windows on it, the separation arguments behind the packing bounds,
the averaging over $\Pi$-orbits with continuous weights and on infinite core paths, the race of the
exponential clocks and the locality and round count are proved on paper only.

\section{Open questions}\label{sec:open}
\begin{itemize}
\item How many rounds does a deterministic $(1+\varepsilon)$-approximation of $\gamma_f$ on planar
  graphs, with no degree bound, need as a function of~$\varepsilon$ (\cite[Open Problem~3]{AS26})?
  By Corollary~\ref{cor:dichotomy}, the
  optimal number of rounds lies between $1/\varepsilon-4$ and $O((c_0/\varepsilon)^{5.83})$, and for
  forests and for outerplanar graphs it is $\Theta(1/\varepsilon)$
  (Theorems~\ref{thm:forests} and~\ref{thm:outerplanar}). Do planar graphs admit local
  windows of radius $\tilde O(1/\varepsilon)$ and weighted boundary $\varepsilon\gamma_f$
  (Section~\ref{sec:forests})? Or is there a planar lower bound that is superlinear in
  $1/\varepsilon$?
\item In a constant number of rounds, deterministic algorithms that output dominating sets cannot
  beat ratio~$3$ on paths~\cite[Thm.~8.7]{AS26}, while for fixed $\varepsilon$ a
  $(1+\varepsilon)$-approximation on planar graphs takes $O(\log^*n)$ rounds~\cite{CHW08}, and
  $\Omega(\log^*n)$ rounds are needed on cycles for $\varepsilon<1/4$~\cite[Prop.~C.1]{ASW26}. How do
  the dependencies on $\varepsilon$ and on $n$ combine, already on paths?
\item What is the worst case of the expected gain of the uniform rule on planar graphs? For the
  best~$p$, it lies between $0.1718$ (Theorem~\ref{thm:planar}) and $0.1741$ (Apollonian networks).
\item Is there a rule in which the centre decisions depend on each other and that provably contracts
  more than $\varphi(p)$ per level? By Proposition~\ref{prop:indep}, rules with independent coins
  cannot go below an exponent of $4.74$ with an analysis that counts only join edges, and by
  Proposition~\ref{prop:jointop} no such analysis of any star contraction goes below $3.81$.
\end{itemize}

\section*{Use of AI}

Claude Opus 5.5 (Anthropic) was used throughout this research: to explore approaches, to develop
and check proofs, to write the simulation scripts and the Lean formalization, and to draft the
text.

\end{document}